\documentclass[%
 reprint,
superscriptaddress,
 amsmath,amssymb,
 aps,
]{revtex4-1}

\usepackage{graphicx}
\usepackage{dcolumn}
\usepackage{bm}
\usepackage{amsfonts}
\usepackage{booktabs}
\usepackage{tabularx}
\usepackage{braket}
\usepackage{amsmath}
\usepackage{siunitx}
\usepackage{soul,color}

\begin{document}


\title{Entanglement swapping between independent and asynchronous integrated photon-pair sources}

\author{Farid Samara}
\author{Nicolas Maring}

\author{Anthony Martin}
\altaffiliation{Currently at Université Côte d’Azur, CNRS, Institut de Physique de Nice (INPHYNI), Parc Valrose, 06108 Nice Cedex 2, France.}

\affiliation{Department of Applied Physics, University of Geneva, Geneva, Switzerland.}
\author{Arslan S. Raja}
\author{Tobias J. Kippenberg}
\affiliation{Institute of Physics, Swiss Federal Institute of Technology Lausanne (EPFL), CH-1015 Lausanne, Switzerland.}
\author{Hugo Zbinden}
\author{Rob Thew}
\email{Robert.Thew@unige.ch}
\affiliation{Department of Applied Physics, University of Geneva, Geneva, Switzerland.}

\date{\today}

\begin{abstract}
Integrated photonics represents a technology that could greatly improve quantum communication networks in terms of cost, size, scaling, and robustness. A key benchmark for this is to demonstrate their performance in complex quantum networking protocols, such as entanglement swapping between independent photon-pair sources. Here, using time-resolved detection, and two independent and integrated Si$_3$N$_4$ microring resonator photon-pair sources, operating in the CW regime at telecom wavelengths, we obtained spectral purities up to $0.97 \pm 0.02$ and a HOM interference visibility between the two sources of $V_{\rm HOM}=93.2 \pm 1.6\,\%$. This results in entanglement swapping visibility as high as $91.2 \pm 3.4\,\%$.
\end{abstract}

\maketitle
\section{Introduction} \label{Introduction}

Entanglement swapping~\cite{Zukowski1993}, where two independent particles can be entangled without ever directly interacting, is a fascinating phenomenon, offering a rich insight into the foundations of quantum physics~\cite{Peres2000,Branciard2010}. Additionally, the protocol is of paramount importance for quantum communication networks, such as quantum repeaters~\cite{Briegel1998,Duan2001,Sangouard2011}, enabling efficient long-distance entanglement distribution. Following the seminal work on entanglement swapping~\cite{Boschi1998,Pan1998}, the protocol has been implemented in a wide array of systems, exploiting entanglement encoded in various degrees of freedom~\cite{deRiedmatten05,PanReview2012,Bernien2013,Rosenfeld2017}. However, these have typically been large, bulky experiments that are not suited to real-world deployment.  

Photonic integrated circuits (PIC) provide a promising solution, in terms of size, scalability, and robustness~\cite{Wang2019}. However, while the performance of integrated photon-pair sources in the telecom regime has advanced in recent years, the demonstration of entanglement swapping between completely independent sources has remained outstanding. Of particular interest for quantum communication, photon-pair sources based on spontaneous four-wave mixing (SFWM) in microring resonators (MRR)~\cite{Grassani2015,Hemsley2016,Ma2017,samara2019,Oser2020} are emerging as a viable technology~\cite{Pasquazi2018,Kues2019}, not only due to their compact size but also their compatibility with standard telecom systems and devices - both the photons and the pump laser are in the telecom regime. In the case of multi-photon experiments, teleportation and entanglement swapping was recently demonstrated using two MRR photon-pair sources~\cite{Llewellyn2019}. However, this was realized with both sources integrated on the same chip and pumped with the same laser, thus not addressing many of the challenges associated with real-world quantum communication. 

A key challenge for distributing entanglement in such scenarios is the synchronization of the sources or the network in general~\cite{Tanzilli2010}. Most of the previous demonstrations of entanglement swapping relied on pulsed photon-pair sources, where pump synchronization is indispensable~\cite{Kaltenbaek2006,Yang2006,Kaltenbaek2009,sun2017,Sun2017_100km}. Alternatively, the combination of continuous-wave (CW) photon-pair sources, and time-resolved detection allows for asynchronous entanglement swapping to be realized~\cite{Zukowski1993,Halder2007}. Time-resolved detection refers to the uncertainty in detection time being smaller than the photons' coherence time. While simplifying the synchronization of sources for entanglement distribution in networks, this approach also overcomes problems due to path length changes and chromatic dispersion in pulsed systems~\cite{sun2017}, facilitating high-quality entanglement distribution over long distances. 

\begin{figure*}
\includegraphics[width=16cm]{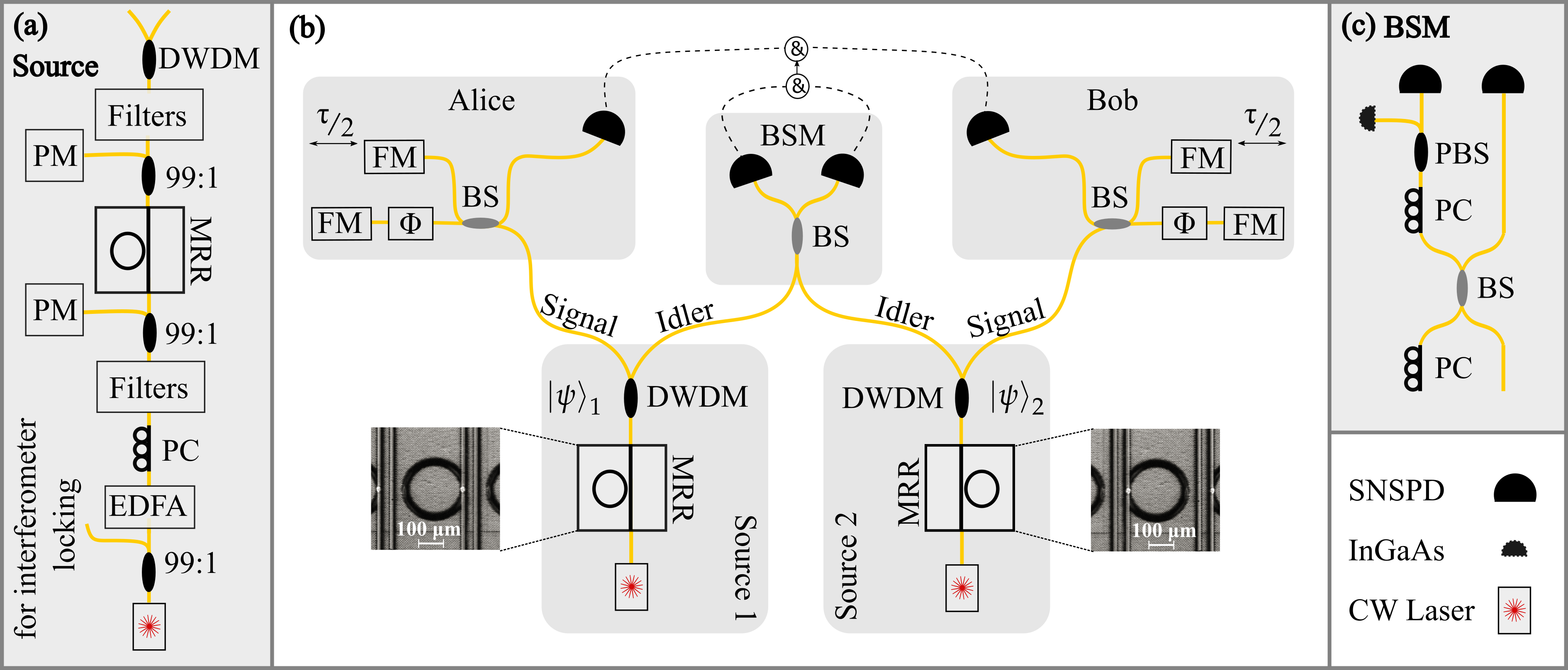}
\caption{Concept and experimental schematic. (a) Photon pair source: A continuous wave (CW) telecom laser is amplified with an erbium-doped fiber amplifier (EDFA). A polarization controller (PC) adjusts the polarization to the TE mode. A tunable filter (before the chip) and a series of commercial dense wavelength division multiplexers (DWDM) reject the EDFA spontaneous emission at the chip input, as well as filter the residual pump and separate signal and idler at the chip output (Filters and DWDM). An optical splitter (99:1) is used to monitor the output pump power via a power meter (PM). (b) Entanglement swapping: Two independent sources prepare energy-time entangled photons. The idler photons are sent to the central node where the Bell-state measurement (BSM) is performed by impinging the photons on a 50:50 beam splitter (BS) and detecting them with a time difference of $\tau = 3.9\,\text{ns}$. The signal photons are distributed to Alice and Bob, where the time-bin entanglement can be verified by interferometers with a temporal imbalance of $\tau$. (c) The setup at the BSM for aligning the polarisation of the two idler photons.}
\label{setup}
\end{figure*}

Central to the entanglement swapping protocol is an interference measurement between one photon from each of the two photon-pair sources, requiring not only a high degree of indistinguishability but also spectral purity~\cite{Osorio2013}. Here, we exploit two independent MRR photon-pair sources, developed on silicon nitride ($\mathrm{Si_3N_4}$) platform, in combination with state-of-the-art, low jitter detectors, to demonstrate high-quality HOM and CW entanglement swapping. The $\mathrm{Si_3N_4}$ based photonic integrated platform has gained a lot of interest in many linear and nonlinear fields due to its ultra-low propagation loss (1 dB/m)~\cite{liu2020high}, wide transparency range, high power handling capability~\cite{gyger2020observation}, CMOS and space compatibility~\cite{Brasch:14}. In particular, $\mathrm{Si_3N_4}$ based soliton microcombs and supercontinuum generation have been already implemented in many system-level applications~\cite{kippenberg2018} and can facilitate integrated quantum optics devices.

\section{The concept}
A schematic of our entanglement swapping implementation is shown in Fig.~\ref{setup}. Two autonomous sources, labeled 1 and 2, independently generate energy-time entangled photon-pairs. The individual photon-pair states belonging to one of the two sources can be written as $\ket{\psi}\propto\sum \ket{t,t}_{si}$, where the subscripts $s$ and $i$ refer to the signal and idler respectively. The combined state of both photon-pairs generated by both sources is given by: 
\begin{equation} \label{eq1}
\begin{split}
\ket{\psi}_{12}&=\ket{\psi}_1\ket{\psi}_2\propto\sum_{t} \ket{t,t}_{si1} \otimes \sum_{t}\ket{t,t}_{si2}\\
= \sum_{t} \sum_{\tau_{p}\geq\tau\geq 0} &\ket{t,t}_{si1}\ket{t+\tau,t+\tau}_{si2}\\
& + \ket{t+\tau,t+\tau}_{si1}\ket{t,t}_{si2}\\
= \sum_{t} \sum_{\tau_{p}\geq\tau\geq 0} &\ket{\Phi^+}_{s}\ket{\Phi^+}_{i}+\ket{\Phi^-}_{s}\ket{\Phi^-}_{i}\\
&+ \ket{\Psi^+}_{s}\ket{\Psi^+}_{i}+\ket{\Psi^-}_{s}\ket{\Psi^-}_{i}
\end{split}
\end{equation}
where $\tau$ is the relative delay between the emission times of the two photon sources, and $\tau_p$ is the pump coherence time. The Bell states in equation~(\ref{eq1}) are defined as: 
\begin{equation} \label{eq2}
\begin{split}
&\ket{\Phi^\pm}_{s(i)} =\ket{t}_{s(i)1}\ket{t}_{s(i)2} \pm \ket{t+\tau}_{s(i)1}\ket{t+\tau}_{s(i)2}\\ &\ket{\Psi^\pm}_{s(i)} =\ket{t}_{s(i)1}\ket{t+\tau}_{s(i)2} \pm \ket{t+\tau}_{s(i)1}\ket{t}_{s(i)2}
\end{split}
\end{equation}
Only one $\tau$ is post-selected in our experiment, facilitating the entanglement swapping analysis with fixed delay interferometers. To guarantee the quantum interference between $\ket{t}$ and $\ket{t+\tau}$ and to avoid single-photon interference, $\tau$ must satisfy the energy-time entanglement condition $\tau_p \gg \tau \gg \tau_c$, where $\tau_c$ is the coherence time of the photons.

After the beam splitter (BS), the two idler photons are projected in one of the four Bell states (equation~(\ref{eq2})). The two Bell states $\ket{\Phi^\pm}_i$ ($\ket{\Psi^\pm}_i$) correspond to both photons in the same (different) time mode (modes). The photons in the states $\ket{\Phi^\pm}_i$ will exit the BS in the same output. In our realization, only the two Bell states $\ket{\Psi^\pm}_i$ can be detected unambiguously. Nevertheless, the detection of the state $\ket{\Psi^+}_i$ is technically challenging, as it requires single-photon detectors with a recovery time shorter than $\tau$. Indeed, after the beam splitter, the (anti-)symmetric state ($\ket{\Psi^-}_i$) $\ket{\Psi^+}_i$ corresponds to both photons exiting in (different) the same spatial (modes) mode. In light of such technical difficulties, and considering the recovery time of our detectors (${\sim}50\,$ns), we consider only the state $\ket{\Psi^-}_i$. As we see in equation~(\ref{eq1}), the detection of the state $\ket{\Psi^-}_{i}$ collapses the other two photons in the state $\ket{\Psi^-}_{s}$. In other words, time-bin entanglement is post-selected from energy-time entanglement without the need for pulsed pumps or their synchronization, greatly facilitating distributed scenarios.

\section{Experimental implementation}
The photon-pair sources use SFWM in $\mathrm{Si_3N_4}$ microring resonators. The microring resonators are fabricated using the photonic Damascene process~\cite{Pfeiffer2016}, enabling devices with low propagation loss ($1\,$\text{dB/m}) waveguides while maintaining anomalous dispersion

Each source, Fig.~\ref{setup}(a), is pumped on resonance by a CW laser (Toptica DL100) amplified by an erbium-doped fiber amplifier (EDFA) in the telecom band. The pump wavelength for source 1 (2) is $1555.86\,$\text{nm} ($1557.43\,$\text{nm}), while the pump power is set to $12.5\,$\text{mW} ($4.5\,$\text{mW}). The chosen pump power values for both sources correspond to a pair generation probability per coherence time ($p$) of 0.006. A tunable bandpass filter and two DWDM filters, reject the amplified spontaneous emission (ASE), giving rise to $135\,\text{dB}$ of pump isolation before the chip. The polarization of the pump light is aligned (using a polarization controller PC) to the quasi-transverse electric (TE)mode, achieving the best input-to-output transmission in the micro-ring resonator (MRR) chip. Lensed fibers are used to couple the light in and out of the $\mathrm{Si_3N_4}$ waveguide. To reject the residual pump at the chip output and separate the signal and idler into different spatial modes, a second filtering block is implemented after the PIC. 

This filtering block consists exclusively of low-loss commercial DWDMs (2 in notch configuration + 2 in passband configuration), which reject the residual pump and direct the signal and idler photons to the appropriate channels. This results in $135\,$dB of pump rejection and $100\,$dB signal-idler isolation. The DWDM's passband of $200\,$GHz is $>$ 400 times larger than the photons' spectra (see Appendix), thus, apart from the insertion losses, no further reduction in the photons flux occurs due to spectral filtering. The average heralding efficiency for the four paths is $13.4\,\%$ (see Table~S2 in Appendix for a complete breakdown of the various contributions). Here the heralding efficiency is defined as the probability of having a photon in the output mode once its herald has been detected. 

A control loop is used to minimize the pump wavelength and MRR resonance detuning by adjusting the temperature of the $\mathrm{Si_3N_4}$ chip, ensuring stable operation for more than 24~hours. The temperature control system is also necessary for aligning the photons' wavelengths within the specific DWDM passband, and to optimize the spectral overlap between the two independent idler photons ($1558.98\,$nm) arriving at the Bell-state measurement (BSM) node. A complete characterization of the photon-pair sources is given in Appendix. 

Once the signal and idler photons are separated, the idler photons are sent to the BSM node consisting of a 50:50 beam splitter (BS), while the signal photons are sent to Alice and Bob (Fig.~\ref{setup}(b)). The photons are detected with homemade superconducting nanowire single-photon detectors (SNSPD), with temporal jitter on the order of $35\,$ps, efficiencies above $82\,\%$ and dark count rates below $500\,$Hz. The arrival time of the photons is registered with a multi-stop time-to-digital converter (IDQ ID900) with bin width and time jitter of $13\,$ps and $20\,$ps respectively. The detection of photons at the BSM detectors, with a relative time difference $\tau = 3.9\,$ns and coincidence window $\tau_w = 52\,$ps, heralds the presence of the signal photons at Alice and Bob. The coincidence window $\tau_w$ was chosen to satisfy the time-resolved detection condition $\tau_c >> \tau_w$, where $\tau_c = 850\,$ps is the average coherence time of both idler photons.

To analyze the entanglement swapping, we use two imbalanced ($\Delta t = \tau$) Franson-type interferometers placed on Alice's and Bob's sides (Fig.~\ref{setup}(b)) that are thermally stabilized and phase-locked (See Appendix). A detection at the BSM at time $t$ and $t + \tau$ projects the original energy-time entanglement of the two sources onto a $\ket{\Psi^-}$ time-bin entangled state shared by Alice and Bob. For this experiment, Alice's and Bob's interferometers are locked with the pump lasers of sources 1 and 2, respectively. Additionally, the frequency difference between the two pump lasers is stabilized by using one of the interferometers as a frequency reference. This ensures that there is no phase drift between Alice's and Bob's interferometers. In a quantum repeater like scenario, the sources could be locked to a relevant phase reference, such as an atomic transition, cavity, or optical frequency comb.

\section{Results}
Spectral purity and indistinguishability of the photons are essential requirements for a successful BSM. Spectral purity can be related to the factorability of the signal-idler joint spectral amplitude or an auto-correlation measurement~\cite{Osorio2013}. Due to the narrow bandwidth of the sources, we use auto-correlation measurements to characterize the spectral purity (see Appendix). The obtained spectral purity for source 1 and 2 are $0.96  \pm 0.01$  and $0.97 \pm 0.02$ respectively, where the ideal value is 1, i.e. for a single-mode thermal source. 

Photon indistinguishability, for all degrees of freedom, is the other essential requirement for a successful BSM~\cite{Hong1987}. Spatial indistinguishability is achieved by using single-mode fibers, while polarization indistinguishability is realized by minimizing the photon flux at the polarizing beam splitter's (PBS) auxiliary output as shown in Fig.~\ref{setup}(c). For the spectral indistinguishably, we chose two MRR devices with comparable Q-factors, thus giving rise to similar photon spectra, with linewidths at full-width half-maximum (FWHM) of 335 and 476\,MHz - The spectral overlap is $98.5\,\%$. Importantly, temporal indistinguishability is ensured by the sub-coherence-time precision, given by a detection resolution of $53\,\text{ps}$ (convolution of the detectors and the multi-stop time-to-digital converter jitters) and a coincidence window of $\tau_w = 52\,$ps. 
\begin{figure}[htbp]
\centering\includegraphics[width=1.0\columnwidth]{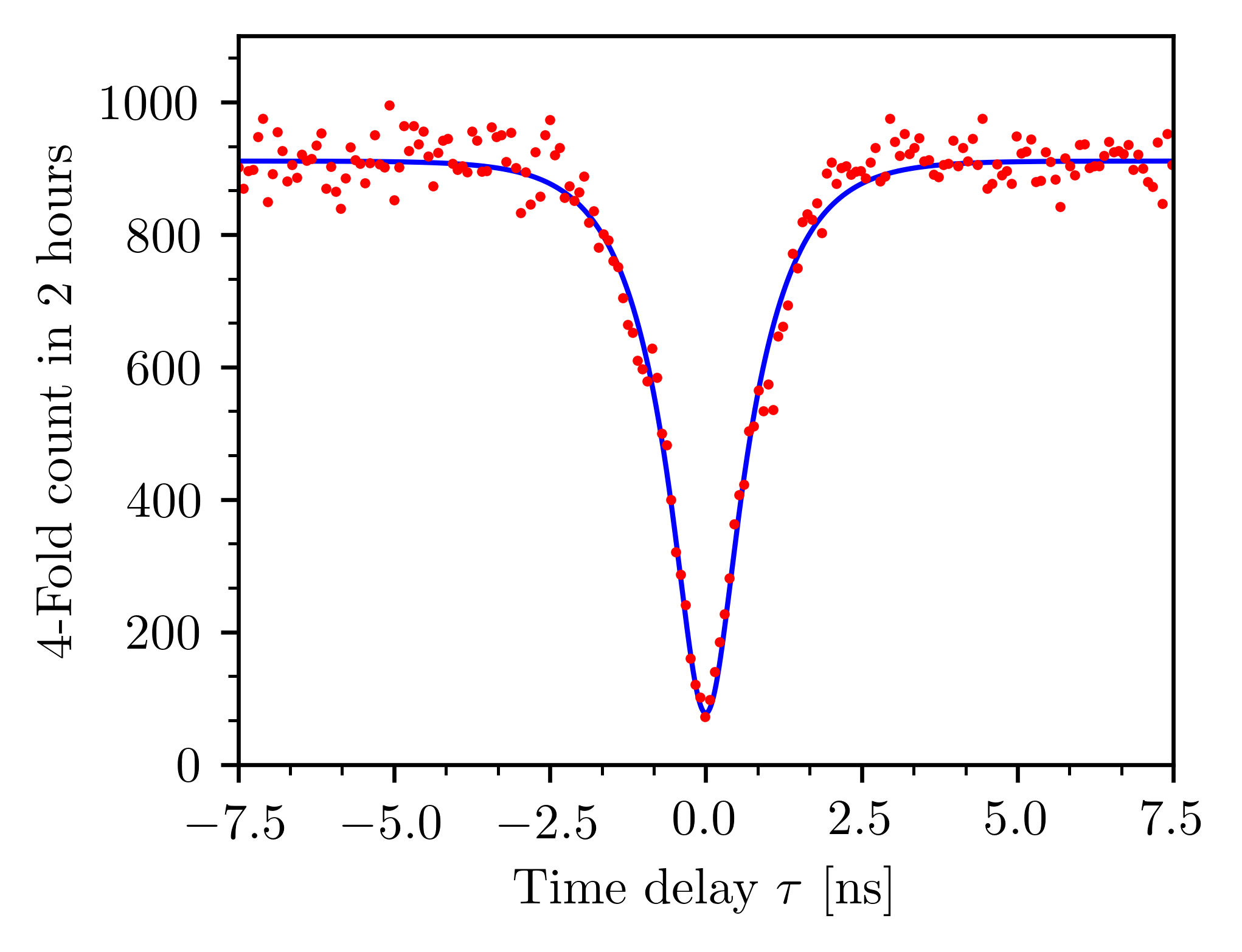}
\caption{Hong–Ou–Mandel (HOM) dip. Four-fold coincidence counts as a function of the time delay $\tau$ between the photons at the central node. The HOM visibility is $93.2 \pm 1.6\,\%$.}.
\label{HOM}
\end{figure}

Photon indistinguishability and spectral purity can be verified by performing a heralded Hong–Ou–Mandel (HOM) measurement~\cite{Hong1987}, where the visibility also provides an operational characterization for the BSM. To perform this measurement, we use the setup illustrated in Fig.~\ref{setup}(b) without the interferometers. The four-fold coincidence histogram as a function of the relative delay between the two photons at the BSM node is shown in Fig.~\ref{HOM}. The data is fitted with a function obtained by the convolution between two double exponential functions describing the temporal shape of the two photons, and the detectors' Gaussian temporal response function. The HOM dip visibility is $ V_{HOM} =93.2 \pm 1.6\,\%$.

\begin{figure}[htbp]
\centering
\includegraphics[width=8.4cm]{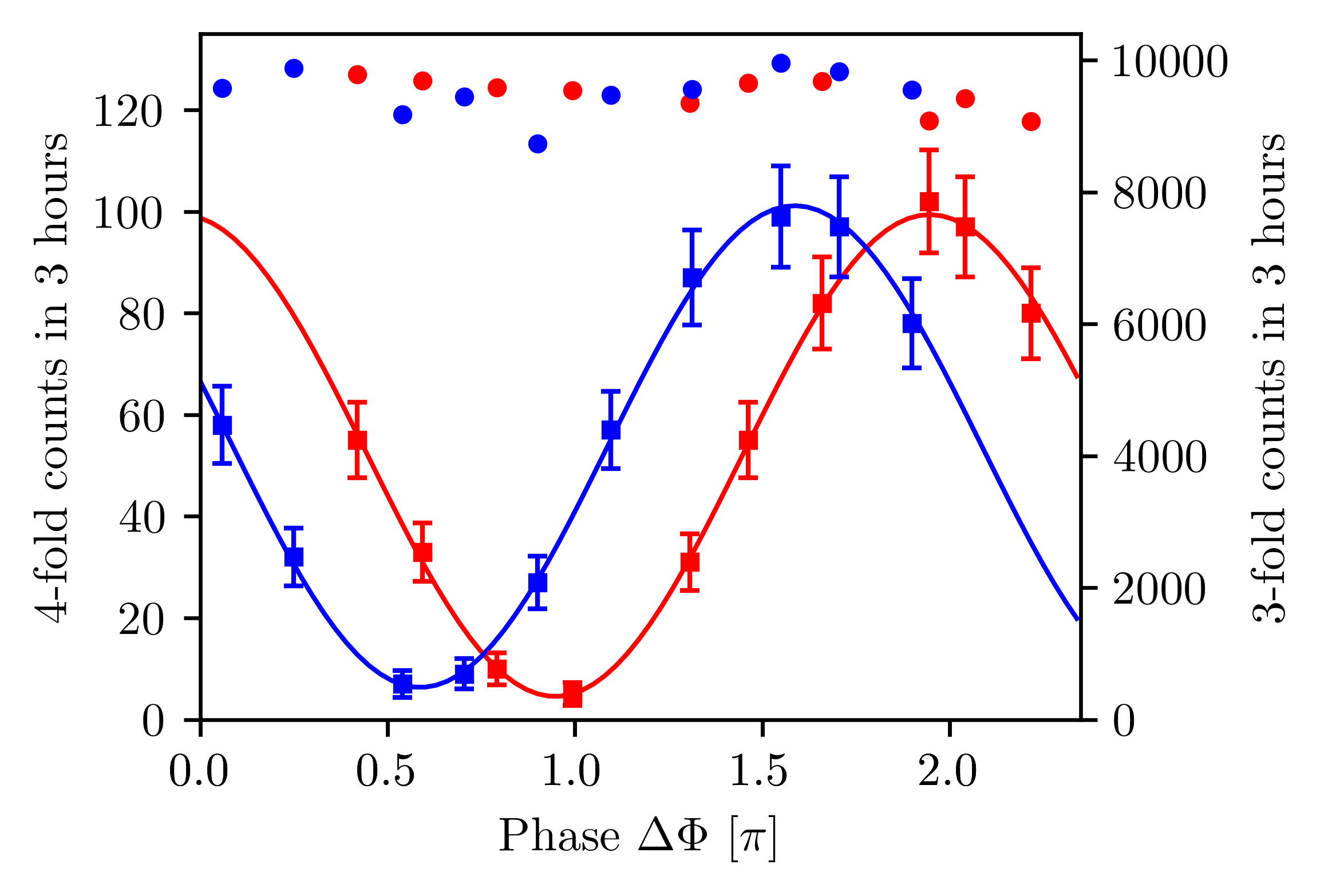}
\caption{Entanglement swapping. (Left axis, square points) Four-fold coincidences as a function of the relative phase $\Delta \Phi$ between the two interferometers. Bob's phase was scanned while Alice's phase was kept constant at two different values corresponding to the two fringes. The visibilities are $V = 88.2 \pm 3.5\,\%$ (blue) and $V = 91.2 \pm 3.4\,\%$ (red). The error bars represent one standard deviation calculated by assuming Poissonian statistics. (Right axis, round points) The corresponding 3-fold coincidences between one detector conditioned on a successful BSM measurement, where no interference is expected or observed.}
\label{BSM}
\end{figure}

To analyze the quality of the entanglement swapping we use the complete setup shown in Fig.~\ref{setup}. Bob's phase is scanned for two different phase settings in Alice's interferometer. Fig.~\ref{BSM} illustrates the four-fold coincidence counts as a function of this phase difference. The interference visibilities of $V_{SWP}=88.2 \pm 3.5\,\%$ and $V_{SWP}=91.2 \pm 3.4\,\%$ clearly indicate a high degree of entanglement. The visibility can also be related to the violation of a CHSH inequality ($S=2\sqrt{2}V$)~\cite{Marcikic2004}, which for the average visibility of $V_{SWP}=89.7 \pm 2.4\,\%$, corresponds to $S=2.54 \pm 0.07 $, allowing us to infer a violation by $7.7$ standard deviations.

\begin{table*}
\caption{Comparison of heralded Hong–Ou–Mandel visibilities ($V_{HOM}$) for pulsed or continuous-wave (CW) regimes, and sources monolithically integrated on the same (On), or separate (Off) chips. In all the studies below photon pairs are generated in the telecom band, except in \cite{Spring2017}, where photon pairs are generated in the visible.}
\centering
\label{Table1}
\begin{tabular}{cccccccccc}
\hline
Ref.              & Source   & Process      &
Pump   & On/Off    & $p$ & $V_{HOM}$  & HOM rate & $V_{Swap.}$ &  Swap. rate   \\
                      &             &   && chip  &     & raw (net)   & [$\times10^3$ cph] & raw    &  cph    \\
\hline
Current           & $\mathrm{Si_3N_4}$ MRR   & SFWM     &CW $@1555/1557\,$nm  & Off & 0.006   & $93\,\%$ ($95\,\%$)  & $0.45$ & $91\,\%$ & $33$ \\
\cite{Llewellyn2019} 2019 & Si MRR      & SFWM    &Pulsed $@1550\,$nm& On&  0.065   & $72\,\%$ ($91\,\%$) & ${\sim}2.25$ &  ${\sim}70\,\%$ & ?\\

\cite{Spring2017} 2017   & Silica WG     & SFWM &Pulsed $@736\,$nm & On&  0.015   & ${\sim}92\,\%$ ($96\,\%$) &  ${\sim}720$ & - & -\\
\cite{Zhang2016}  2016     & SOI nanowire  & SFWM &Pulsed $@1555/1556\,$nm  & Off& 0.02  & $88\,\%$ (-) &${\sim} 0.023$ & - & -     \\ 
\cite{Harada2011}   2011   & Si WG       & SFWM  &Pulsed $@1550\,$nm & Off& 0.025 & $73\,\%$ (-) & ${\sim} 0.04$ & - &  - \\ 
\cite{Tanzilli2010} 2010   & PPLN WG      & SPDC  &Pulsed $@768\,$nm & Off& 0.04 & $93\,\%$ ($99\,\%$) & ${\sim} 0.01$ & - & -\\ 
\cite{Halder2007}   2007   & PPLN WG      & SPDC  &CW  $@780\,$nm     & Off& 0.02  & $77\,\%$ (-) & ${\sim}0.001$ & $63\,\%$   & 5\\ 
\hline
  \end{tabular}
\end{table*}

The imperfect HOM and entanglement swapping visibilities can be attributed to the following contributions: double pairs ($99.0\,\%$), photonic noise ($98.0\,\%$), and imperfect spectral purity ($96.5\,\%$). For entanglement swapping, one must also consider the non-ideal energy-time entangled states (see Appendix) generated by source~1 ($98.7\,\%$) and source~2 ($98.8\,\%$). The expected HOM and entanglement swapping visibilities of $V_{HOM}^{(th.)}=93.6\,\%$ and $V_{SWP}^{(th.)}=91.3\,\%$ are given by multiplying the aforementioned contributions, and they are in a good agreement with the experimentally observed values of $V_{HOM}=93.2~\pm~1.6\,\%$ and $V_{SWP}=89.7 \pm 2.4\,\%$.

For the above visibilities estimations, the fluctuations and drift ($400\,\text{MHz}$ in 10 hours) of the spectral overlap have a negligible effect as it is less than the spectral uncertainty given by the detection (${\sim}20\,\text{GHz}$). The effect of the polarization fluctuations and detector dark counts can also be neglected. The photonic noise is mainly being generated by the $\mathrm{Si_3N_4}$ chip (see Appendix and~\cite{samara2019}), its effect was estimated by modelling the noise as detector dark counts~\cite{Sekatski2012} . The timing jitter of the SNSPD ($35\,\text{ps}$) and multi-stop time-to-digital converter ($20\,\text{ps}$) gives a finite time resolution of $53\,\text{ps}$, the effect of which is reflected by the imperfect spectral purity.

\section{Discussion } 
\label{Discussion}

Firstly, it is important to point out that the maximum achievable HOM visibilities for MRR sources were theorized to be fundamentally limited to ${\sim} 92\,\%$, given by the maximum attainable spectral purity of the basic circuit design~\cite{Helt2010}. However, Huang~{\it et al.}~\cite{Huang2010} have shown that time-resolved, single-mode, detection can achieve high spectral purity even if the photons are initially in a non-factorable state. The single-mode detection condition can be quantitatively stated as $BT<4$, where $B$ is the photons bandwidth, and $T$ is the temporal resolution of the photon detection scheme. In our case $BT\sim0.02$ (on average $B = 387\,\text{MHz}$ and $T = 53\,\text{ps}$). By exploiting time-resolved detection we overcome this limit and achieve spectral purities up to $97\,\%$ from a standard MRR design.

The rates here are, in general, very good, although there are again fundamental limits associated with the circuit design that could be further improved with more complex approaches~\cite{Tison2017}. This approach would allow one to optimise the coupling between the resonator and pump independently of the resonator and pairs~\cite{Vernon2016}. Currently, the efficiency of coupling out of the cavity and into the waveguide is only $38\,\%$. The overall heralding efficiency could be further improved with advanced 2D taper designs~\cite{Liu18} to improve the coupling into the fiber.

A significant limiting factor for the current scheme is the generation of noise photons~\cite{samara2019}. This forces us to operate at lower pump powers and hence lower photon pair generation probabilities, thus sacrificing the overall four-fold rates. Further research to better understand and develop solutions to resolve this noise issue are essential to exploit the potential of the $\mathrm{Si_3N_4}$ PIC technology. This would allow further optimization of the trade-off between the four-fold rate and visibility, as a function of the detection coincidence window $\tau_w$ (see Appendix). 

In Table~\ref{Table1} we compare some of the key parameters for a range of integrated photonic systems. Our results significantly improve over the most conceptually similar experiment~\cite{Halder2007}, which also exploited time-resolved detection, as well as some pulsed experiments in SOI nanowires~\cite{Zhang2016} and Si waveguides~\cite{Harada2011} for both rates and visibilities. The results for PPLN waveguides~\cite{Tanzilli2010}, which have provided a benchmark for some time, give comparable performance in both visibilities and rates - if one factors out the low efficiency detectors used in that experiment. An interesting case is for Silica waveguides~\cite{Spring2017}, which, while not at telecom wavelengths, avoid the problem of excess loss due to filtering of the latter case by engineering pure photons, achieving visibilities on a par with the current work, but with significantly higher rates. 

The most straightforward comparison is with the recent results for MRR in Si~\cite{Llewellyn2019}, where the advantage of time-resolved detection can be clearly seen by the significantly improved visibility. The rates for these two are not so different, although we note that the photon pair probability for this work was an order of magnitude lower. Indeed, we are currently limited by noise generated in the devices, but in principle, we could increase our rates a hundredfold without detrimental effects on the visibility (see Appendix), which would surpass most of these systems and be on a par with~\cite{Spring2017} and more traditional bulk sources engineered to produce pure photons~\cite{Bruno2014}.

\section{Conclusion }

We have demonstrated an asynchronous entanglement swapping experiment exploiting $\mathrm{Si_3N_4}$ MRR photon-pair sources and time-resolved detection. The time-resolved detection allowed us to surpass what were previously thought to be fundamental limits for purity and HOM interference in basic MRR schemes. This consequently allowed us to achieve a record interference visibility of over $91\,\%$ for entanglement swapping between truly independent PIC-based sources. This could be further improved with a better understanding and identification of solutions to deal with the spurious noise sources inherent in many integrated photonic sources. Similarly, four-fold rates could be dramatically improved through a more complex circuit design. Nonetheless, the $\mathrm{Si_3N_4}$ MRR have clearly demonstrated their potential for integration in more complex and distributed quantum network architectures and protocols. 
\section*{Funding}
This work was supported by the Swiss National Science Foundation SNSF, Grant No. 200020\_182664, Grant No. 176563 (BRIDGE)and the NCCR QSIT. This material is based upon work supported by the Air Force Office of Scientific Research, Air Force Materiel Command, USAF under Award No. FA9550-19-1-0250.
\begin{acknowledgments}
The authors would like to thank M.~Caloz and M.~Karpov for development of the SNSPD and the MRR, and M.~Afzelius, P.~Caspar, A.~Boaron for useful discussions and technical support.
\end{acknowledgments}

\appendix

\section{Photon-pair sources characterisation}
The microring resonator (MRR) photon-pair sources are fabricated on the $\mathrm{Si_3N_4}$ platform. The two sources are selected from different fabrication batches, they were chosen based on reasonably similar resonance characteristics around $1550\,$nm. The resonance parameters of both sources, including full width at half maximum of the photons' Lorentzian spectrum $\Delta v$, the coherence time $\tau_c$ (defined as $1/{\pi\Delta v}$), the Q-factor, the free spectral range FSR, and the central wavelengths $\lambda$ are all summarized in Table~\ref{Table2}.
\begin{figure}[htbp]
\centering
\includegraphics[width=8.4cm]{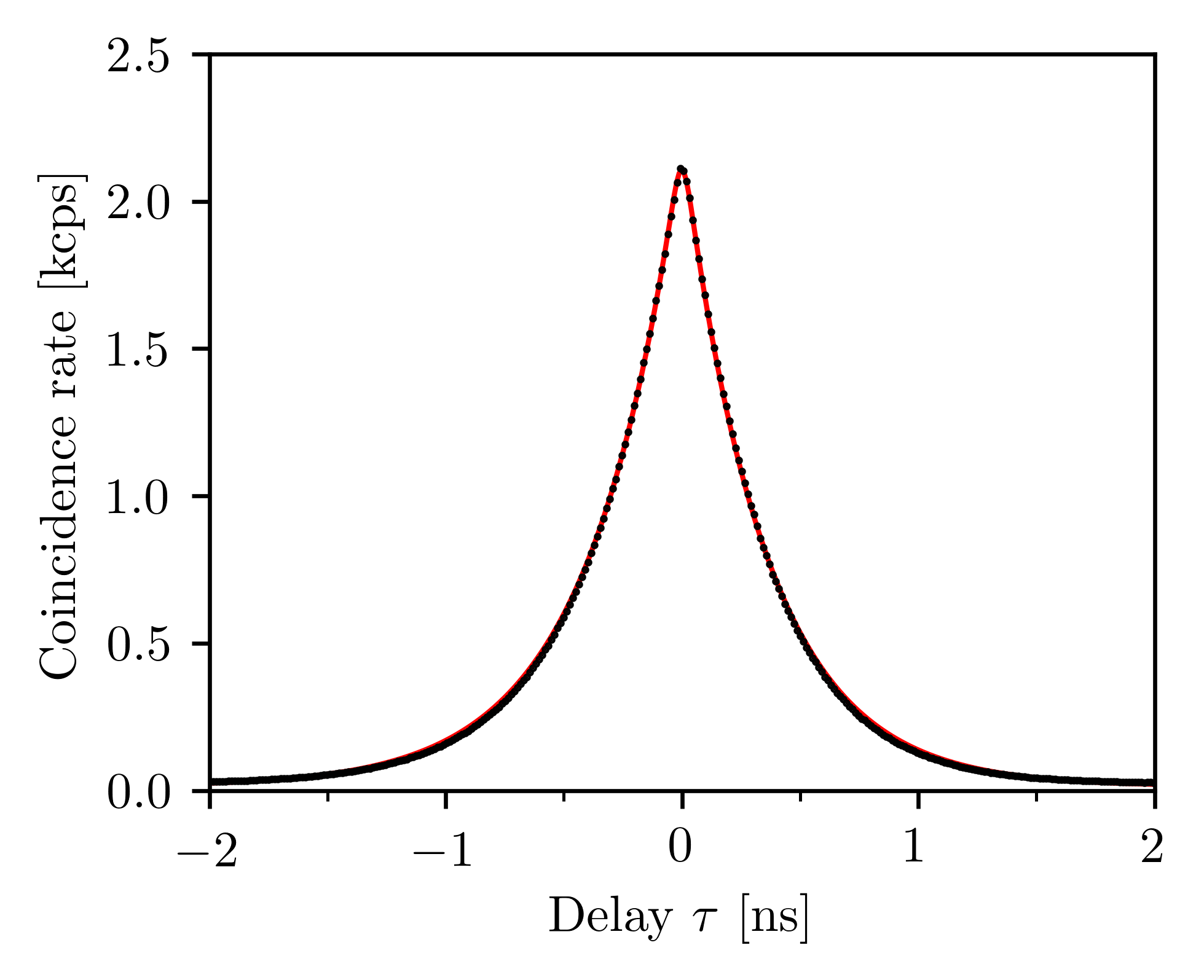}
\caption{Signal-idler coincidence histogram of source 2 when pumped with $12.5\,\text{mW}$. Histogram resolution is $13\,\text{ps}$. Fitting function is explained in the text.}
\label{Cross}
\end{figure}

To characterize the sources individually, we first perform a cross-correlation analysis. Figure~\ref{Cross} reports an example of a coincidence histogram between the signal and idler of source~2 obtained with a pump power of $12.5\,\text{mW}$. Here, the experimental data is fitted with a function given by an asymmetric double exponential, convoluted with a Gaussian distribution~\cite{Clausen2014}. The asymmetric double exponential part accounts for the temporal shape of the photons, including the slight miss-match between signal and idler resonances, and the Gaussian part accounts for the detectors' timing jitter. From such a fitting, we can extract the coherence time and spectral width of our signal and idler photons individually (Table~S1).
\begin{table*}
\caption{\bf Photon-pair sources characterisation. When both signal and idler data are provided, the later is given in the parentheses.}
\centering
\label{Table2}
\begin{tabular}{c|ccccccccccc}
\hline
&$\Delta v$ & $\tau_{c}$ & Q-factor& FSR & $\lambda_s (\lambda_i)$&  $\lambda_p$ & PGR &${\eta}_{H}$ & $g^{(2)}(0)$\\
&[MHz]  & [ps] && [nm]  & [nm]  & [nm] & [$\text{mW}^{-2}]$ &$[\%]$ & \\
\hline
Source 1  & 476 (437)   & 669 (728) &$440 (401)\times10^3$  &1.557& 1552.75 (1558.98)& 1555.86 & $55 \times 10^3$ & 10.5 (13.6)  & 1.96   \\
Source 2  & 335 (301)   & 950 (1057) &$640 (579)\times10^3$& 1.554&  1555.87 (1558.98)& 1557.43 &$300 \times 10^3$  & 13.0 (16.4) & 1.97   \\
\hline
  \end{tabular}
\end{table*}
\begin{figure}[htbp]
     \centering
     \includegraphics[width=8.4cm]{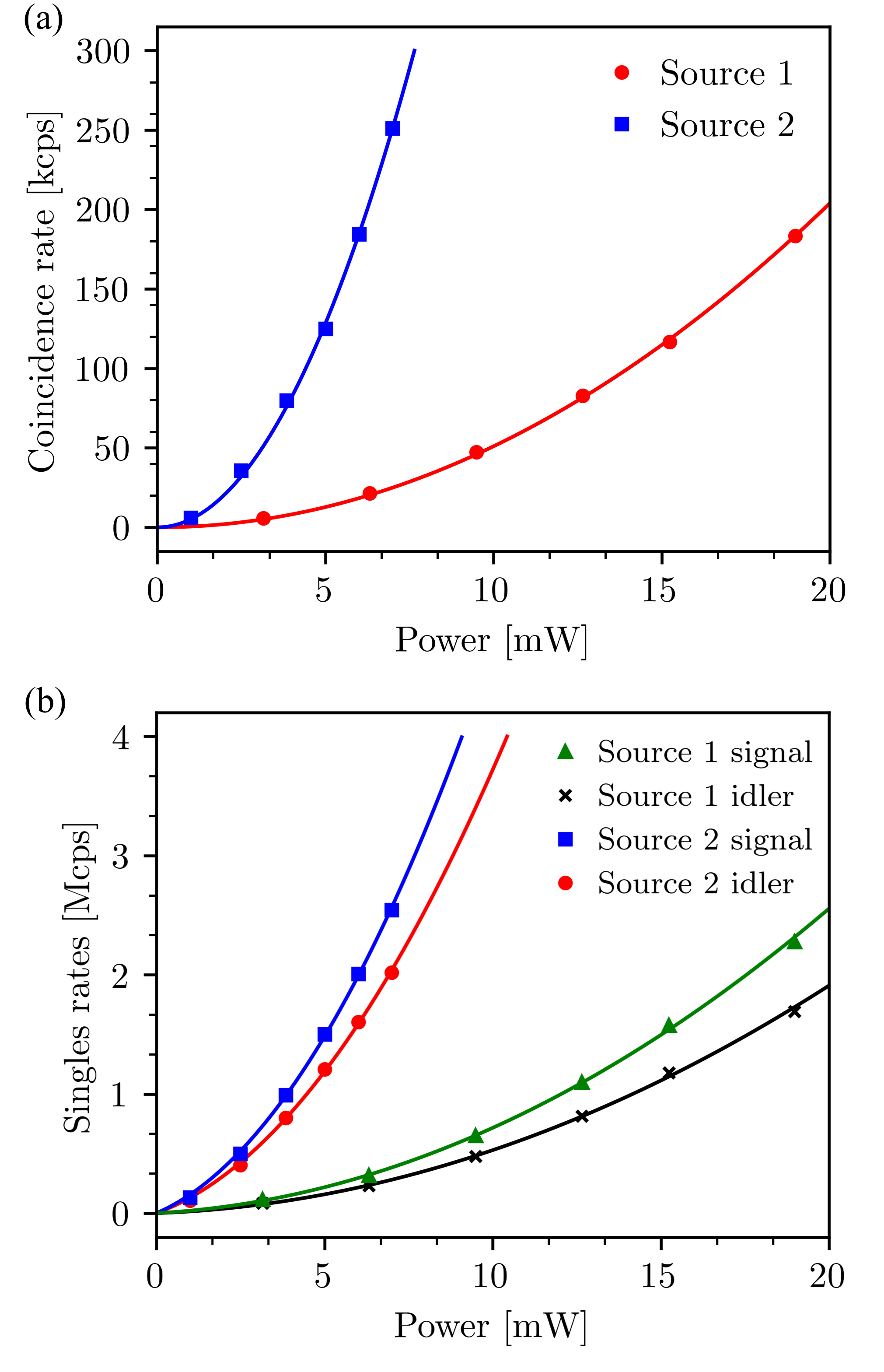}
     \caption{(a) Signal-idler coincidences rate as a function of the pump power. The experimental points are given by integrating the total counts in the un-normalised cross-correlation measurements. (b) Singles rates as a function of the pump power. The experimental data are fitted with $aP+bP^2$, where P is the pump power, $a$ is the linear photonic noise term and $b$ is the quadratic SFWM term.}
      \label{S2}
\end{figure}
The coincidences and singles as a function of the pump power are reported in Figure~\ref{S2}. Here, the coincidences are taken as the total coincidences inside the correlation peak, minus the accidental coincidences outside the peak; while the singles are taken as the singles on-resonance, minus the singles off-resonance. As confirmed from the fitting, the coincidences scale quadratically with the pump power, in accordance with what is expected from the spontaneous four-wave mixing (SFWM) process. Nevertheless, in addition to the quadratic power dependency, a linear component is observed in the singles versus power relation~\cite{samara2019}. Such a linear component is not expected to be the product of the SFWM process, giving rise to uncorrelated photons, which for us, acts as a photonic noise. As the singles in Figure~\ref{S2}(b), the photonic noise that we are referring to here is generated on-resonance inside the MRR chip. By considering only the singles that are coming from the SFWM process, i.e. the quadratic term in the fitting equation, the pair generation rate PGR and the heralding efficiency ${\eta}_{H}$ can be calculated (see Table~\ref{Table1}). The difference between the PGR of the two sources can be explained by the difference between the coupling regimes (coupling between the cavity and the waveguide), also notable from different Q-factors.

\begin{figure}[htbp]
\centering
\includegraphics[width=8.4cm]{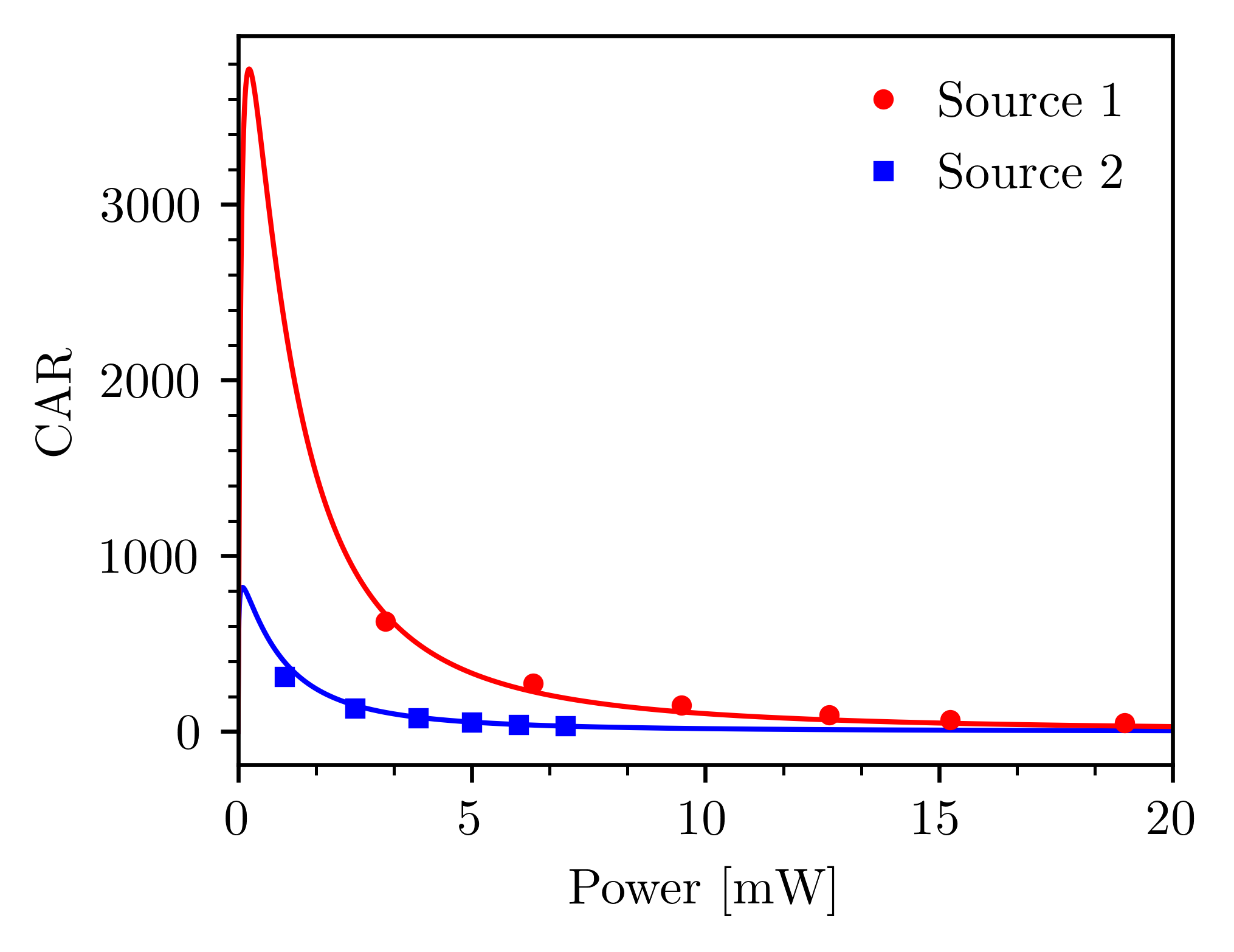}
\caption{Coincidence-to-accidental ratio (CAR) as a function of the pump power. See text for explanation. }
\label{CAR}
\end{figure}

The coincidences-to-accidental ratio (CAR) gives a measure of the ratio between the correlated photon-pairs and the unwanted coincidences noise. The CAR as a function of the pump power is reported in Figure~\ref{CAR}. The experimental data is fitted with a model that takes into account the photonic noise and the detectors dark counts. At high pump power, the CAR is limited by the double pair contribution, while at low pump power the limiting factor is the detector dark counts. 

\begin{table}[htbp]
 \caption{\bf Breakdown of losses. When both signal and idler data are provided, the later is given in the parentheses.}
 \centering
 \begin{tabular}{c|cc}
 \label{Table3}
  & Source 1 & Source 2\\
 \hline
 EL : Cavity extraction [dB]    & 5.2       & 3.3 \\
 CL : Waveguide-to-fiber [dB]    & 2.0       & 3.0 \\
 FL : Fiber components [dB]    & 1.8 (2.1) & 1.8 (2.3) \\
 \hline
 Total [dB] & 9.2       & 8.4   \\
   \end{tabular}
\end{table}
Table~\ref{Table3} reports a breakdown of the transmission losses for each of the four paths in the HOM experiment. The coupling losses CL, and the fiber component transmission losses FL are measured directly by a power transmission measurement. Here the fiber components are the combination of power monitors, DWDMs for pump rejection and signal idler isolation, and polarization controllers (used to align the polarization for maximizing the SNSPDs detection efficiencies). For the idler paths, we consider both outputs of the 50:50 beam splitter BS, thus only its insertion losses are considered. This, together with the polarization aligning setup (Figure 2(c) main text), explains the slight difference between the signal and idler transmission losses.

Finally, photons need to couple out of the MRR cavity and into the bus waveguide, which accounts for additional transmission losses which we call extraction losses EL. These EL can be inferred by subtracting the CL and FL from the total transmission losses calculated from the heralding efficiencies ${\eta}_{H}$ (Table~\ref{Table1}). For the entanglement swapping experiment, the additional interferometers in the idler paths and the DWDMs that are required for their phase-locking give on average an additional $2.1\,\text{dB}$ and $0.7\,\text{dB}$ of transmission losses in each path. 

\begin{figure}[htbp]
\centering
\includegraphics[width=8.4cm]{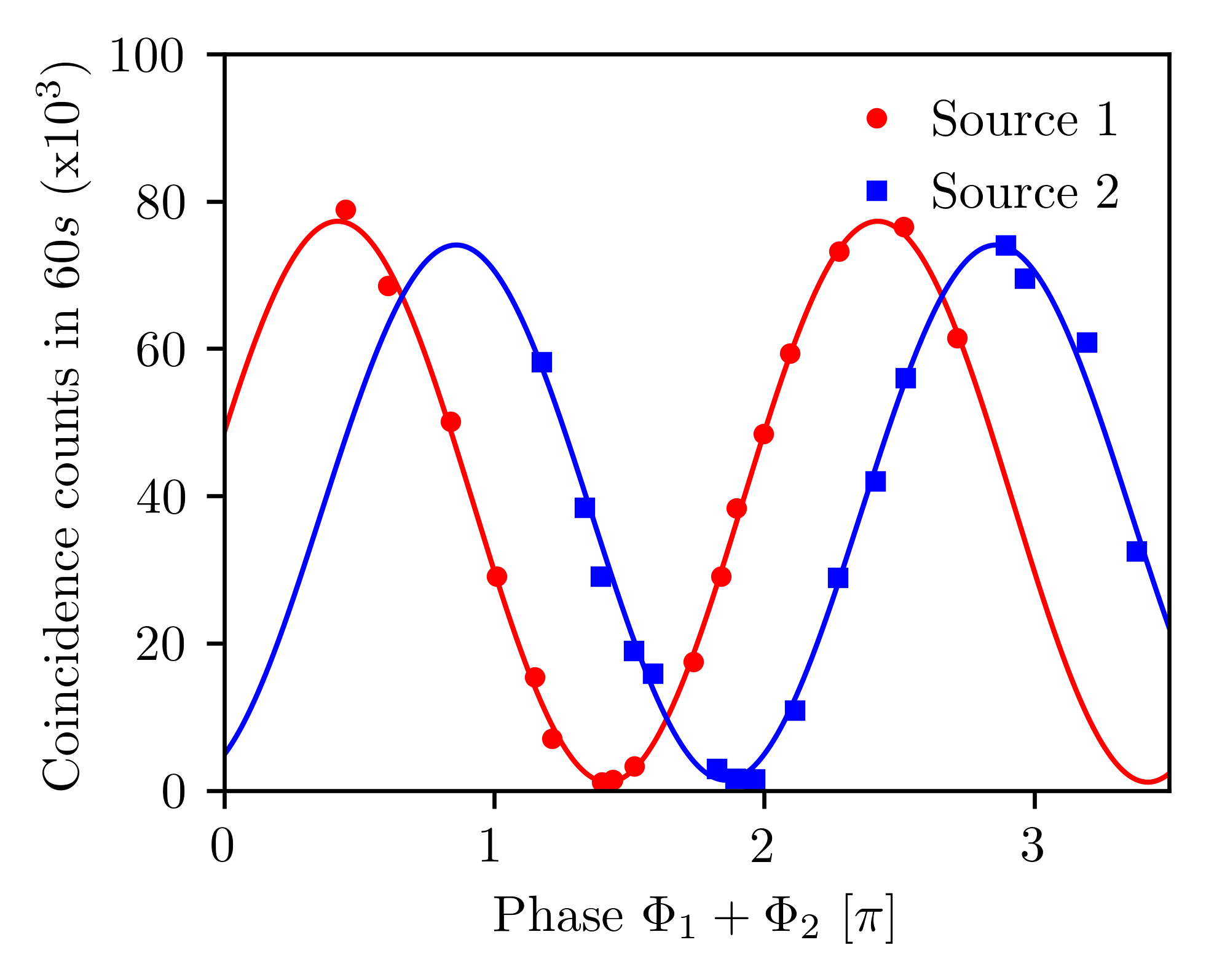}
\caption{Energy-time entanglement. Coincidence counts of the interfering term between the signal and idler as a function of the phase sum $\Phi_1 + \Phi_2 $ of the two analysing interferometers. The relative shift between the two fringes is induced for the sake of representation. Accidental coincidences are not subtracted. }
\label{ET}
\end{figure}

The sources emit energy-time entangled photons. The HOM and entanglement swapping visibilities are directly related to the quality of the energy-time entanglement. The quality of the energy-time entanglement can be verified individually for each source by passing the correlated photon-pairs in two imbalanced interferometers and observing the interfering term in the signal idler coincidences histogram as a function of the sum of the two interferometers phases $\Phi_1 + \Phi_2$ \cite{Franson1989}.  Figure~\ref{ET} reports the coincidence fringes using two interferometers with path-length differences equal to $\tau =3.9\,\text{ns}$. The raw (net) visibilities are $97.1 \pm 1.1\,\%$ ($98.7 \pm 1\,\%$) and $96.3 \pm 2.1\,\%$ ( $98.8 \pm 2.1\,\%$) for source~1 and~2 respectively.

\section{Spectral purity}
High spectral purity is crucial for high visibility quantum interferences between independent sources. Spectrally pure photons mean that their joint spectral amplitude JSA is uncorrelated; the detection of the heralding photon does not change the spectral distribution of the heralded photon. 

There are several methods for characterizing the spectral purity of a given photon-pair source~\cite{Zielnicki2018}. Here, due to the narrow spectral shape of the photons, we use the unheralded auto-correlation measurement $g^{(2)}(\tau)$. For single-mode thermal statistics, we expect a $g^{(2)}(0) = 2$, while for a Poissonian superposition of several thermal modes (multimode), the  $g^{(2)}(0)$ tends to~1. The spectral purity $\mathcal{P}$, and the number of Schmidt modes K can be subsequently derived as $g^{(2)}(0) = 1+\mathcal{P} = 1 + 1/K$. The obtained $g^{(2)}(0)$ for source~1~(2) is $1.96 \pm 0.01$ ($1.97 \pm 0.02$), with a corresponding $\mathcal{P}$ and K values of $0.96 \pm 0.01$ ($0.97 \pm 0.02$) and $1.04 \pm 0.01$ ($1.03 \pm 0.02$).

\begin{figure}[htbp]
\centering
\includegraphics[width=8.4cm]{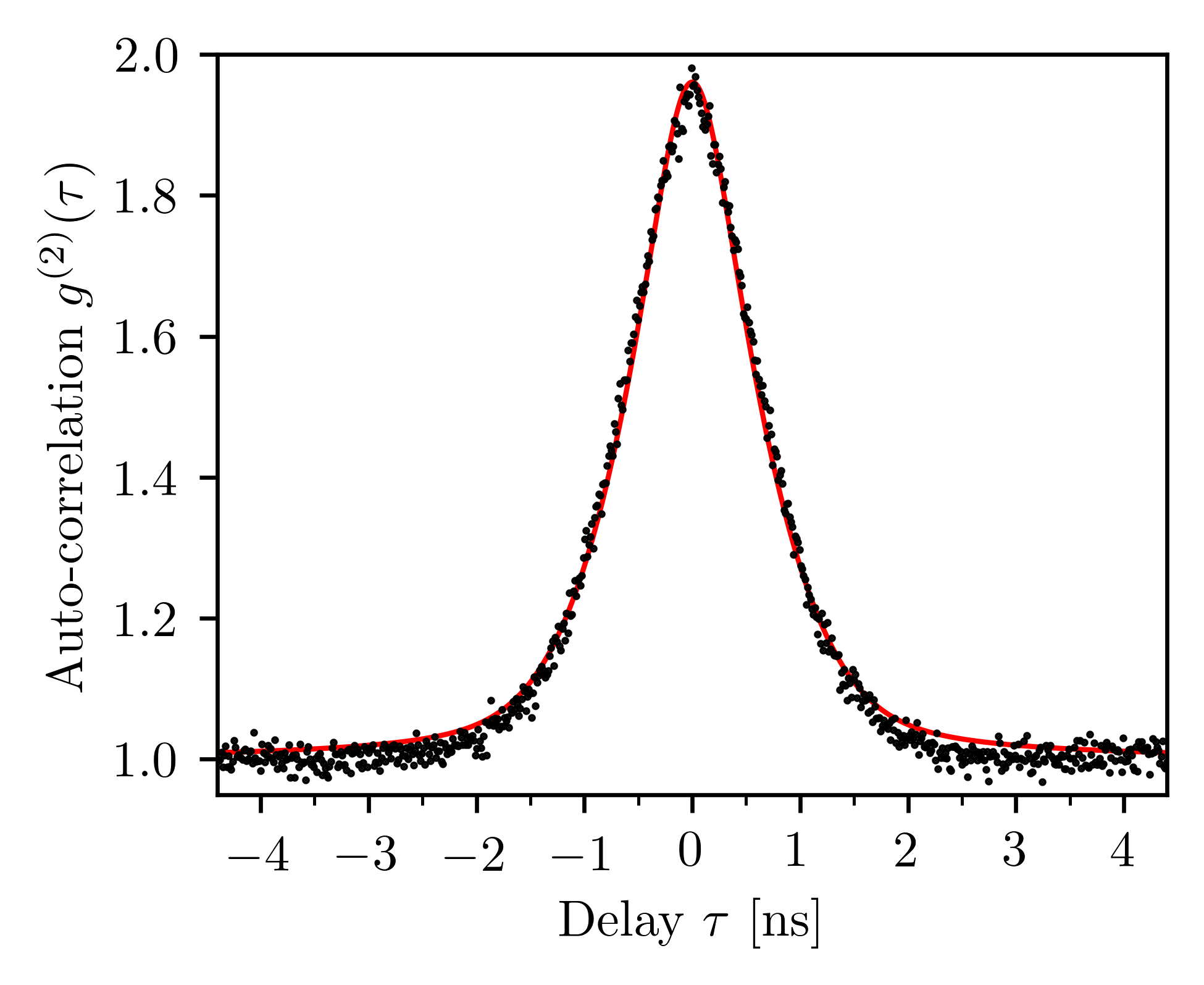}
\caption{Unheralded auto-correlation measurement. Measurement performed on the idler mode of source 1.}
\label{g2}
\end{figure}

Figure~\ref{g2} reports an example of an auto-correlation measurement. Here, the experimental data is fitted with a Pseudo-Voigt function, thus accounting for the Lorentzian shape of the two-photon component~\cite{Luo_2015,Luo_2017} and the Gaussian jitter of the superconducting nanowire single-photon detectors SNSPDs.

\section{Spectral indistinguishability}
\label{Stabilisation}
Spectral indistinguishability is another important element that must be guaranteed for high visibility quantum interferences. In this work, spectral indistinguishability is guaranteed, firstly, by aligning the central wavelengths of both idler photons, and secondly by detecting with a small time resolution ($53\,\text{ps}$), projecting the photons with ${\sim}20\,\text{GHz}$ spectral uncertainty, i.e. about 42 to 66 times larger than the spectral width of the photons.

\begin{figure}[htbp]
\centering
\includegraphics[width=8.4cm]{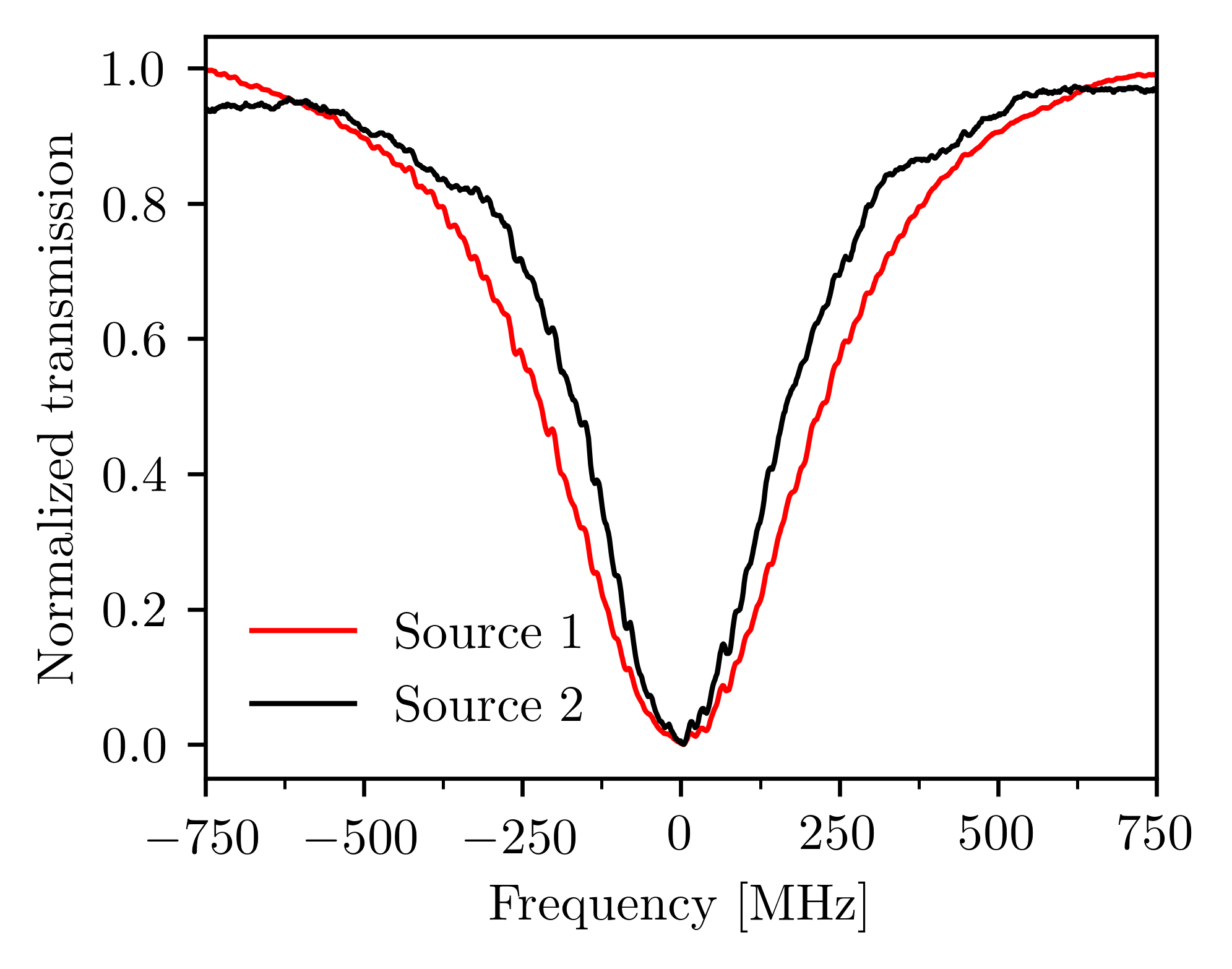}
\caption{Power transmission scan of the two idler photon' resonances. The FWHMs of the Lorentzian shaped resonances $\Delta v$ are $437$ and $301\,\text{MHz}$ for source~1 and~2 respectively. The overlap integral is $98.48\,\%$.   }
\label{SpectrumAlignment1}
\end{figure}

The spectrum alignment of the idler photons is achieved by temperature tuning of the chip. A Peltier controls the temperature of each PIC through a feedback system. At our working wavelengths, the temperature coefficient is $2.74\,\text{MHz.mK}^{-1}$. By setting the temperatures of the PICs to $T_1=308.65\,\text{K}$ and $T_2=309.85\,\text{K}$, the spectra of both idler photons were aligned between them, and in the center of the idler DWDM's bandwidth (ITU channel N.23, $200\,\text{GHz}$). Table~\ref{Table1} reports the operating wavelengths that are obtained after performing the spectral alignment. Figure~\ref{SpectrumAlignment1} reports a transmission scan of the aligned idlers' resonances. The obtained spectral overlap is $98.48\,\%$, limited by a slight miss-match between the spectral width as shown in Table~\ref{Table1}. 

The central frequency of each laser drifts with time, leading to pump-resonance decoupling. By locking the chip temperature to the resonance dip, we avoid the pump-resonance decoupling, achieving on-resonance operation for more than 24~hours. This is achieved by a computer algorithm that monitors the residual pump power at the PIC output and generates an error signal that acts on the set point of the temperature controller system. 

\begin{figure}[!h]
\centering
\includegraphics[width=8.4cm]{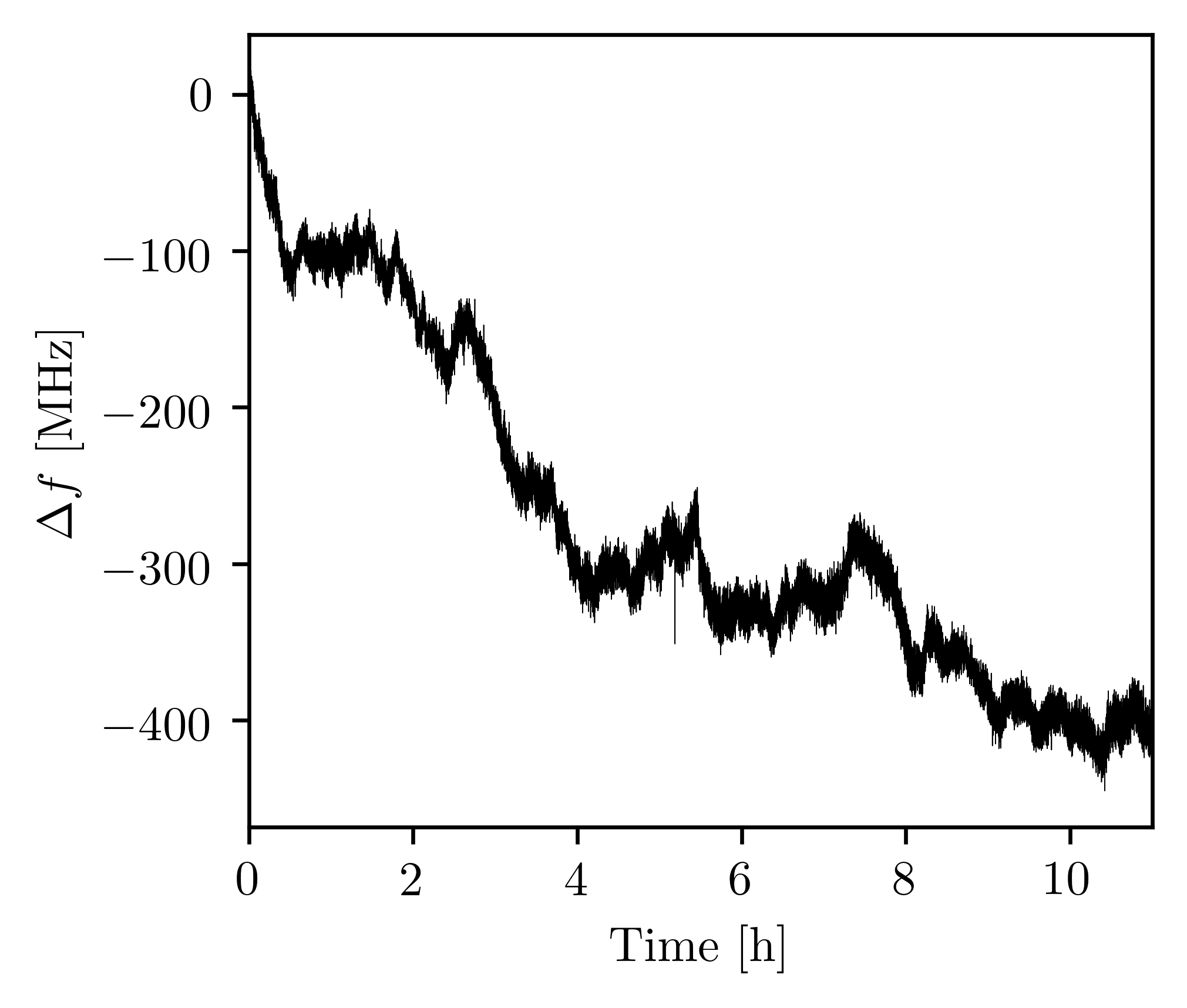}
\caption{Pump lasers' drift in time. The two central frequencies of the independent pump lasers are set to the same value at $t = 0$, then the miss-match $\Delta f$ is monitored as a function of time.}
\label{CentralFrequencies}
\end{figure}

The two MRRs are pumped with two different CW pump lasers (Toptica DL100). The lasers' drift induces a central frequency miss-match $\Delta f$ as shown in Figure~\ref{CentralFrequencies}, causing a spectral overlap degradation. For the HOM measurement, nothing further is done to compensate for the time-dependant spectral overlap degradation. Indeed, as confirmed from Figure~\ref{CentralFrequencies}, the effect of the relative drift between the two lasers over the relatively short period of the HOM measurement (2~hours) can be neglected since the drift is much shorter than the spectral uncertainty given by the fast detection (${\sim}20\,\text{GHz}$). 

\section{Interferometers stabilisation scheme}
For the entanglement swapping experiment, the drift of the lasers frequencies produces an unwanted drift in the individual and relative phases of the interferometers. This is exacerbated by the long measurement time (30 hours) required to acquire each fringe in Figure~3. To compensate for this, we implement a locking system to stabilise the absolute phase of each interferometer $\Phi_1$ and $\Phi_2$, in addition to the relative phase $\Delta \Phi$. This is done by locking Alice's (Bob's) interferometer on the laser pump of source~1~(2). For this aim, a fraction of pump power before the PIC -where power fluctuation is minimum- is extracted, and consequently re-injected into the corresponding interferometer. The error signal is then used to act on the interferometer's piezo, thus keeping a constant phase for the passing photons. To compensate for the relative frequency drift between the two lasers, part of source~2 laser pump power was also injected in Alice's interferometer, thus giving an error signal on which the frequency of the pump laser is locked.

\section{Four-fold rate and visibility}
\begin{figure}[htbp]
\centering
\includegraphics[width=8.4cm]{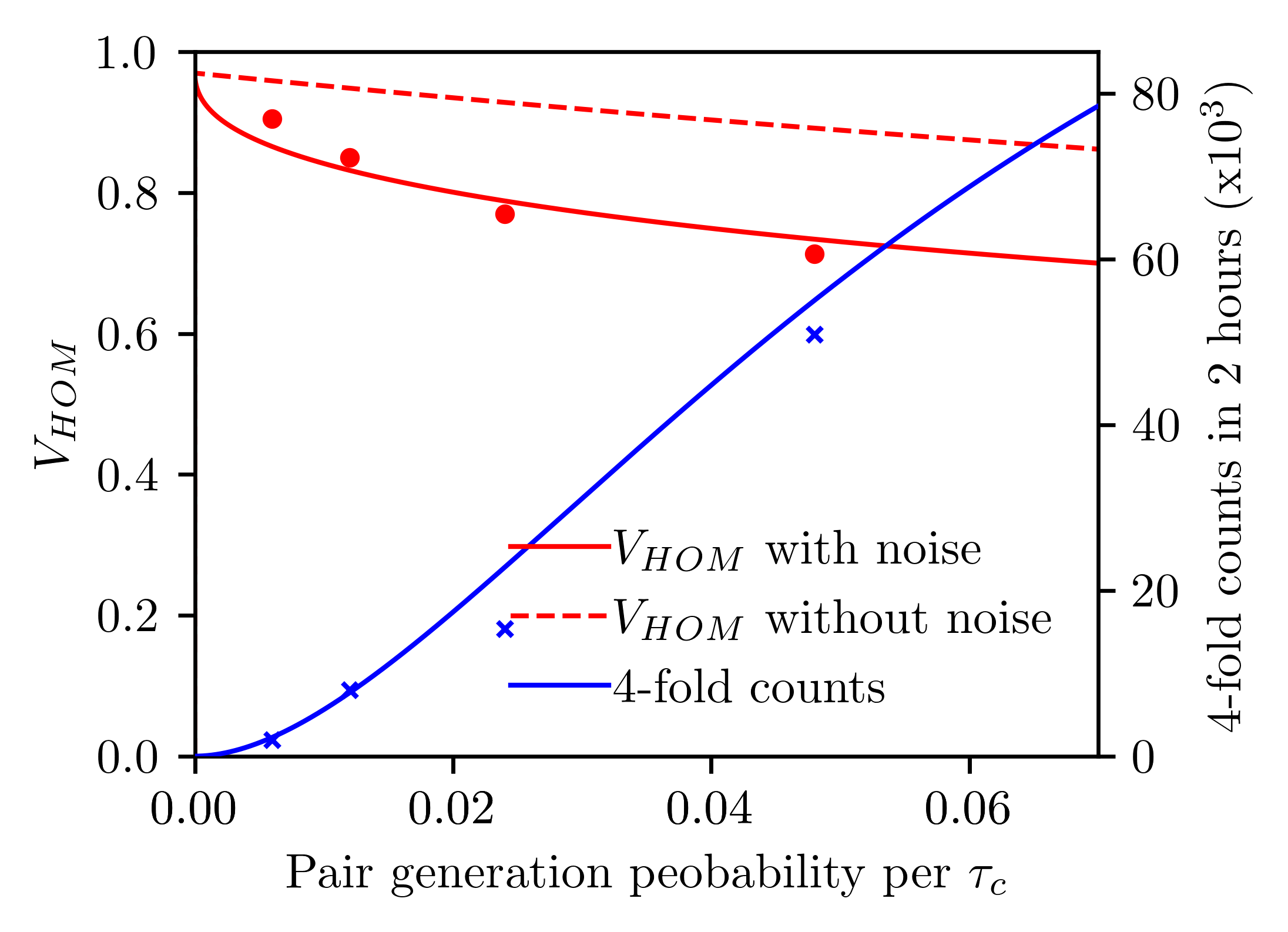}
\caption{Four-fold counts and $V_{HOM}$ as a function of the pair generation probability per coherence time. To show the effect of the photonic noise on the $V_{HOM}$, the experimental points are supported with two models \cite{Sekatski2012} that differ between them by including or excluding the observed photonic noise. The experimental data of the four-fold rate was fitted with a model that include the effect of SNSPDs efficiency saturation at high count rates. This explains the non-linear scaling of the four-fold rate.}
\label{HOMvsPower}
\end{figure}

The average four-fold rates in our HOM and entanglement swapping experiments are about 450 and 33~cph (max fringe) respectively. One can increase the four-fold rate by increasing the pair generation probability, but at the cost of lower HOM and entanglement swapping visibilities due to double-pair contribution. The four-fold rate and $V_{HOM}$ trade-off is reported in Figure~\ref{HOMvsPower}. While it is possible to improve the four-fold rate by working at higher pair generation probabilities (fixed at 0.006 per coherence time in the present work), the $V_{HOM}$ reduction is clearly faster than what is expected  for a photon pair source without the presence of photonic noise. This motivates further study into the origin of the photonic noise with the hope of its eventual mitigation.

\begin{figure}[htbp]
\centering
\includegraphics[width=8.4cm]{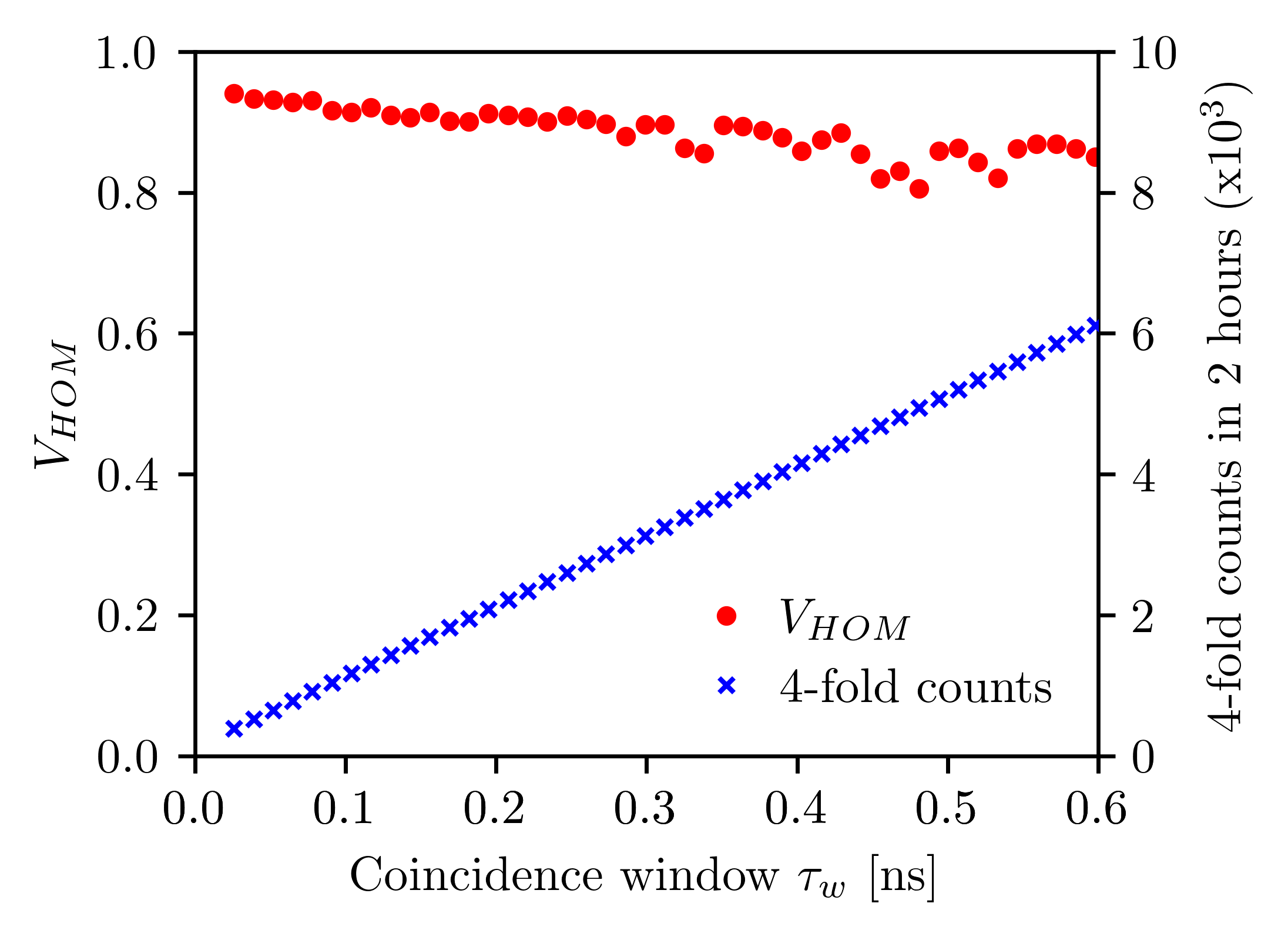}
\caption{Four-fold counts and $V_{HOM}$ as a function of the coincidence window $\tau_w$.}
\label{HOMvsWindow}
\end{figure}

A second factor that is directly affecting the four-fold rate is the coincidence window $\tau_w$ at the BSM node. In the present work, the BSM coincidence window is set to $52\,\text{ps}$, guaranteeing the time-resolving condition $\tau_c >> \tau_w$, therefore resulting in high HOM and entanglement swapping visibilities. Note that a shorter window will not necessarily mean better visibilities since the chosen value is already comparable with our detection scheme resolution. However, depending on the application at hand, one can chose to elongate the coincidence window, thus achieving higher four-fold rates at the cost of lower visibilities. The trade-off between four-fold rate and HOM visibility as a function of the coincidence window is reported in Figure~\ref{HOMvsWindow}.

\bibliography{Bibliography_file}

\begin{thebibliography}{51}%
\makeatletter
\providecommand \@ifxundefined [1]{%
 \@ifx{#1\undefined}
}%
\providecommand \@ifnum [1]{%
 \ifnum #1\expandafter \@firstoftwo
 \else \expandafter \@secondoftwo
 \fi
}%
\providecommand \@ifx [1]{%
 \ifx #1\expandafter \@firstoftwo
 \else \expandafter \@secondoftwo
 \fi
}%
\providecommand \natexlab [1]{#1}%
\providecommand \enquote  [1]{``#1''}%
\providecommand \bibnamefont  [1]{#1}%
\providecommand \bibfnamefont [1]{#1}%
\providecommand \citenamefont [1]{#1}%
\providecommand \href@noop [0]{\@secondoftwo}%
\providecommand \href [0]{\begingroup \@sanitize@url \@href}%
\providecommand \@href[1]{\@@startlink{#1}\@@href}%
\providecommand \@@href[1]{\endgroup#1\@@endlink}%
\providecommand \@sanitize@url [0]{\catcode `\\12\catcode `\$12\catcode
  `\&12\catcode `\#12\catcode `\^12\catcode `\_12\catcode `\%12\relax}%
\providecommand \@@startlink[1]{}%
\providecommand \@@endlink[0]{}%
\providecommand \url  [0]{\begingroup\@sanitize@url \@url }%
\providecommand \@url [1]{\endgroup\@href {#1}{\urlprefix }}%
\providecommand \urlprefix  [0]{URL }%
\providecommand \Eprint [0]{\href }%
\providecommand \doibase [0]{http://dx.doi.org/}%
\providecommand \selectlanguage [0]{\@gobble}%
\providecommand \bibinfo  [0]{\@secondoftwo}%
\providecommand \bibfield  [0]{\@secondoftwo}%
\providecommand \translation [1]{[#1]}%
\providecommand \BibitemOpen [0]{}%
\providecommand \bibitemStop [0]{}%
\providecommand \bibitemNoStop [0]{.\EOS\space}%
\providecommand \EOS [0]{\spacefactor3000\relax}%
\providecommand \BibitemShut  [1]{\csname bibitem#1\endcsname}%
\let\auto@bib@innerbib\@empty
\bibitem [{\citenamefont {\ifmmode~\dot{Z}\else
  \.{Z}\fi{}ukowski}(1993)}]{Zukowski1993}%
  \BibitemOpen
  \bibfield  {author} {\bibinfo {author} {\bibfnamefont {M.}~\bibnamefont
  {\ifmmode~\dot{Z}\else \.{Z}\fi{}ukowski}},\ }\href {\doibase
  10.1103/physrevlett.71.4287} {\bibfield  {journal} {\bibinfo  {journal}
  {Physical Review Letters}\ }\textbf {\bibinfo {volume} {71}},\ \bibinfo
  {pages} {4287} (\bibinfo {year} {1993})}\BibitemShut {NoStop}%
\bibitem [{\citenamefont {Peres}(2000)}]{Peres2000}%
  \BibitemOpen
  \bibfield  {author} {\bibinfo {author} {\bibfnamefont {A.}~\bibnamefont
  {Peres}},\ }\href {\doibase 10.1080/09500340008244032} {\bibfield  {journal}
  {\bibinfo  {journal} {Journal of Modern Optics}\ }\textbf {\bibinfo {volume}
  {47}},\ \bibinfo {pages} {139} (\bibinfo {year} {2000})}\BibitemShut
  {NoStop}%
\bibitem [{\citenamefont {Branciard}\ \emph {et~al.}(2010)\citenamefont
  {Branciard}, \citenamefont {Gisin},\ and\ \citenamefont
  {Pironio}}]{Branciard2010}%
  \BibitemOpen
  \bibfield  {author} {\bibinfo {author} {\bibfnamefont {C.}~\bibnamefont
  {Branciard}}, \bibinfo {author} {\bibfnamefont {N.}~\bibnamefont {Gisin}}, \
  and\ \bibinfo {author} {\bibfnamefont {S.}~\bibnamefont {Pironio}},\ }\href
  {\doibase 10.1103/PhysRevLett.104.170401} {\bibfield  {journal} {\bibinfo
  {journal} {Phys. Rev. Lett.}\ }\textbf {\bibinfo {volume} {104}},\ \bibinfo
  {pages} {170401} (\bibinfo {year} {2010})}\BibitemShut {NoStop}%
\bibitem [{\citenamefont {Briegel}\ \emph {et~al.}(1998)\citenamefont
  {Briegel}, \citenamefont {D\"{u}r}, \citenamefont {Cirac},\ and\
  \citenamefont {Zoller}}]{Briegel1998}%
  \BibitemOpen
  \bibfield  {author} {\bibinfo {author} {\bibfnamefont {H.-J.}\ \bibnamefont
  {Briegel}}, \bibinfo {author} {\bibfnamefont {W.}~\bibnamefont {D\"{u}r}},
  \bibinfo {author} {\bibfnamefont {J.~I.}\ \bibnamefont {Cirac}}, \ and\
  \bibinfo {author} {\bibfnamefont {P.}~\bibnamefont {Zoller}},\ }\href
  {\doibase 10.1103/physrevlett.81.5932} {\bibfield  {journal} {\bibinfo
  {journal} {Physical Review Letters}\ }\textbf {\bibinfo {volume} {81}},\
  \bibinfo {pages} {5932} (\bibinfo {year} {1998})}\BibitemShut {NoStop}%
\bibitem [{\citenamefont {Duan}\ \emph {et~al.}(2001)\citenamefont {Duan},
  \citenamefont {Lukin}, \citenamefont {Cirac},\ and\ \citenamefont
  {Zoller}}]{Duan2001}%
  \BibitemOpen
  \bibfield  {author} {\bibinfo {author} {\bibfnamefont {L.~M.}\ \bibnamefont
  {Duan}}, \bibinfo {author} {\bibfnamefont {M.~D.}\ \bibnamefont {Lukin}},
  \bibinfo {author} {\bibfnamefont {J.~I.}\ \bibnamefont {Cirac}}, \ and\
  \bibinfo {author} {\bibfnamefont {P.}~\bibnamefont {Zoller}},\ }\href
  {\doibase 10.1038/35106500} {\bibfield  {journal} {\bibinfo  {journal}
  {Nature}\ }\textbf {\bibinfo {volume} {414}},\ \bibinfo {pages} {413}
  (\bibinfo {year} {2001})},\ \Eprint {http://arxiv.org/abs/quant-ph/0105105}
  {quant-ph/0105105} \BibitemShut {NoStop}%
\bibitem [{\citenamefont {Sangouard}\ \emph {et~al.}(2011)\citenamefont
  {Sangouard}, \citenamefont {Simon}, \citenamefont {Riedmatten},\ and\
  \citenamefont {Gisin}}]{Sangouard2011}%
  \BibitemOpen
  \bibfield  {author} {\bibinfo {author} {\bibfnamefont {N.}~\bibnamefont
  {Sangouard}}, \bibinfo {author} {\bibfnamefont {C.}~\bibnamefont {Simon}},
  \bibinfo {author} {\bibfnamefont {H.~d.}\ \bibnamefont {Riedmatten}}, \ and\
  \bibinfo {author} {\bibfnamefont {N.}~\bibnamefont {Gisin}},\ }\href
  {\doibase 10.1103/revmodphys.83.33} {\bibfield  {journal} {\bibinfo
  {journal} {Reviews of Modern Physics}\ }\textbf {\bibinfo {volume} {83}},\
  \bibinfo {pages} {33} (\bibinfo {year} {2011})}\BibitemShut {NoStop}%
\bibitem [{\citenamefont {Boschi}\ \emph {et~al.}(1998)\citenamefont {Boschi},
  \citenamefont {Branca}, \citenamefont {Martini}, \citenamefont {Hardy},\ and\
  \citenamefont {Popescu}}]{Boschi1998}%
  \BibitemOpen
  \bibfield  {author} {\bibinfo {author} {\bibfnamefont {D.}~\bibnamefont
  {Boschi}}, \bibinfo {author} {\bibfnamefont {S.}~\bibnamefont {Branca}},
  \bibinfo {author} {\bibfnamefont {F.~D.}\ \bibnamefont {Martini}}, \bibinfo
  {author} {\bibfnamefont {L.}~\bibnamefont {Hardy}}, \ and\ \bibinfo {author}
  {\bibfnamefont {S.}~\bibnamefont {Popescu}},\ }\href {\doibase
  10.1103/PhysRevLett.80.1121} {\bibfield  {journal} {\bibinfo  {journal}
  {Physical Review Letters}\ }\textbf {\bibinfo {volume} {80}},\ \bibinfo
  {pages} {1121} (\bibinfo {year} {1998})}\BibitemShut {NoStop}%
\bibitem [{\citenamefont {Pan}\ \emph {et~al.}(1998)\citenamefont {Pan},
  \citenamefont {Bouwmeester}, \citenamefont {Weinfurter},\ and\ \citenamefont
  {Zeilinger}}]{Pan1998}%
  \BibitemOpen
  \bibfield  {author} {\bibinfo {author} {\bibfnamefont {J.-W.}\ \bibnamefont
  {Pan}}, \bibinfo {author} {\bibfnamefont {D.}~\bibnamefont {Bouwmeester}},
  \bibinfo {author} {\bibfnamefont {H.}~\bibnamefont {Weinfurter}}, \ and\
  \bibinfo {author} {\bibfnamefont {A.}~\bibnamefont {Zeilinger}},\ }\href
  {\doibase 10.1103/PhysRevLett.80.3891} {\bibfield  {journal} {\bibinfo
  {journal} {Physical Review Letters}\ }\textbf {\bibinfo {volume} {80}},\
  \bibinfo {pages} {3891} (\bibinfo {year} {1998})}\BibitemShut {NoStop}%
\bibitem [{\citenamefont {de~Riedmatten}\ \emph {et~al.}(2005)\citenamefont
  {de~Riedmatten}, \citenamefont {Marcikic}, \citenamefont {van Houwelingen},
  \citenamefont {Tittel}, \citenamefont {Zbinden},\ and\ \citenamefont
  {Gisin}}]{deRiedmatten05}%
  \BibitemOpen
  \bibfield  {author} {\bibinfo {author} {\bibfnamefont {H.}~\bibnamefont
  {de~Riedmatten}}, \bibinfo {author} {\bibfnamefont {I.}~\bibnamefont
  {Marcikic}}, \bibinfo {author} {\bibfnamefont {J.~A.~W.}\ \bibnamefont {van
  Houwelingen}}, \bibinfo {author} {\bibfnamefont {W.}~\bibnamefont {Tittel}},
  \bibinfo {author} {\bibfnamefont {H.}~\bibnamefont {Zbinden}}, \ and\
  \bibinfo {author} {\bibfnamefont {N.}~\bibnamefont {Gisin}},\ }\href
  {\doibase 10.1103/PhysRevA.71.050302} {\bibfield  {journal} {\bibinfo
  {journal} {Phys. Rev. A}\ }\textbf {\bibinfo {volume} {71}},\ \bibinfo
  {pages} {050302} (\bibinfo {year} {2005})}\BibitemShut {NoStop}%
\bibitem [{\citenamefont {Pan}\ \emph {et~al.}(2012)\citenamefont {Pan},
  \citenamefont {Chen}, \citenamefont {Lu}, \citenamefont {Weinfurter},
  \citenamefont {Zeilinger},\ and\ \citenamefont {\ifmmode~\dot{Z}\else
  \.{Z}\fi{}ukowski}}]{PanReview2012}%
  \BibitemOpen
  \bibfield  {author} {\bibinfo {author} {\bibfnamefont {J.-W.}\ \bibnamefont
  {Pan}}, \bibinfo {author} {\bibfnamefont {Z.-B.}\ \bibnamefont {Chen}},
  \bibinfo {author} {\bibfnamefont {C.-Y.}\ \bibnamefont {Lu}}, \bibinfo
  {author} {\bibfnamefont {H.}~\bibnamefont {Weinfurter}}, \bibinfo {author}
  {\bibfnamefont {A.}~\bibnamefont {Zeilinger}}, \ and\ \bibinfo {author}
  {\bibfnamefont {M.}~\bibnamefont {\ifmmode~\dot{Z}\else \.{Z}\fi{}ukowski}},\
  }\href {\doibase 10.1103/revmodphys.84.777} {\bibfield  {journal} {\bibinfo
  {journal} {Reviews of Modern Physics}\ }\textbf {\bibinfo {volume} {84}},\
  \bibinfo {pages} {777} (\bibinfo {year} {2012})},\ \Eprint
  {http://arxiv.org/abs/0805.2853} {0805.2853} \BibitemShut {NoStop}%
\bibitem [{\citenamefont {Bernien}\ \emph {et~al.}(2013)\citenamefont
  {Bernien}, \citenamefont {Hensen}, \citenamefont {Pfaff}, \citenamefont
  {Koolstra}, \citenamefont {Blok}, \citenamefont {Robledo}, \citenamefont
  {Taminiau}, \citenamefont {Markham}, \citenamefont {Twitchen}, \citenamefont
  {Childress},\ and\ \citenamefont {Hanson}}]{Bernien2013}%
  \BibitemOpen
  \bibfield  {author} {\bibinfo {author} {\bibfnamefont {H.}~\bibnamefont
  {Bernien}}, \bibinfo {author} {\bibfnamefont {B.}~\bibnamefont {Hensen}},
  \bibinfo {author} {\bibfnamefont {W.}~\bibnamefont {Pfaff}}, \bibinfo
  {author} {\bibfnamefont {G.}~\bibnamefont {Koolstra}}, \bibinfo {author}
  {\bibfnamefont {M.~S.}\ \bibnamefont {Blok}}, \bibinfo {author}
  {\bibfnamefont {L.}~\bibnamefont {Robledo}}, \bibinfo {author} {\bibfnamefont
  {T.~H.}\ \bibnamefont {Taminiau}}, \bibinfo {author} {\bibfnamefont
  {M.}~\bibnamefont {Markham}}, \bibinfo {author} {\bibfnamefont {D.~J.}\
  \bibnamefont {Twitchen}}, \bibinfo {author} {\bibfnamefont {L.}~\bibnamefont
  {Childress}}, \ and\ \bibinfo {author} {\bibfnamefont {R.}~\bibnamefont
  {Hanson}},\ }\href {\doibase 10.1038/nature12016} {\bibfield  {journal}
  {\bibinfo  {journal} {Nature}\ }\textbf {\bibinfo {volume} {497}},\ \bibinfo
  {pages} {86} (\bibinfo {year} {2013})}\BibitemShut {NoStop}%
\bibitem [{\citenamefont {Rosenfeld}\ \emph {et~al.}(2017)\citenamefont
  {Rosenfeld}, \citenamefont {Burchardt}, \citenamefont {Garthoff},
  \citenamefont {Redeker}, \citenamefont {Ortegel}, \citenamefont {Rau},\ and\
  \citenamefont {Weinfurter}}]{Rosenfeld2017}%
  \BibitemOpen
  \bibfield  {author} {\bibinfo {author} {\bibfnamefont {W.}~\bibnamefont
  {Rosenfeld}}, \bibinfo {author} {\bibfnamefont {D.}~\bibnamefont
  {Burchardt}}, \bibinfo {author} {\bibfnamefont {R.}~\bibnamefont {Garthoff}},
  \bibinfo {author} {\bibfnamefont {K.}~\bibnamefont {Redeker}}, \bibinfo
  {author} {\bibfnamefont {N.}~\bibnamefont {Ortegel}}, \bibinfo {author}
  {\bibfnamefont {M.}~\bibnamefont {Rau}}, \ and\ \bibinfo {author}
  {\bibfnamefont {H.}~\bibnamefont {Weinfurter}},\ }\href {\doibase
  10.1103/physrevlett.119.010402} {\bibfield  {journal} {\bibinfo  {journal}
  {Physical Review Letters}\ }\textbf {\bibinfo {volume} {119}},\ \bibinfo
  {pages} {010402} (\bibinfo {year} {2017})}\BibitemShut {NoStop}%
\bibitem [{\citenamefont {Wang}\ \emph {et~al.}(2019)\citenamefont {Wang},
  \citenamefont {Sciarrino}, \citenamefont {Laing},\ and\ \citenamefont
  {Thompson}}]{Wang2019}%
  \BibitemOpen
  \bibfield  {author} {\bibinfo {author} {\bibfnamefont {J.}~\bibnamefont
  {Wang}}, \bibinfo {author} {\bibfnamefont {F.}~\bibnamefont {Sciarrino}},
  \bibinfo {author} {\bibfnamefont {A.}~\bibnamefont {Laing}}, \ and\ \bibinfo
  {author} {\bibfnamefont {M.~G.}\ \bibnamefont {Thompson}},\ }\href {\doibase
  10.1038/s41566-019-0532-1} {\bibfield  {journal} {\bibinfo  {journal} {Nature
  Photonics}\ }\textbf {\bibinfo {volume} {14}},\ \bibinfo {pages} {273}
  (\bibinfo {year} {2019})},\ \Eprint {http://arxiv.org/abs/2005.01948}
  {2005.01948} \BibitemShut {NoStop}%
\bibitem [{\citenamefont {Grassani}\ \emph {et~al.}(2015)\citenamefont
  {Grassani}, \citenamefont {Azzini}, \citenamefont {Liscidini}, \citenamefont
  {Galli}, \citenamefont {Strain}, \citenamefont {Sorel}, \citenamefont
  {Sipe},\ and\ \citenamefont {Bajoni}}]{Grassani2015}%
  \BibitemOpen
  \bibfield  {author} {\bibinfo {author} {\bibfnamefont {D.}~\bibnamefont
  {Grassani}}, \bibinfo {author} {\bibfnamefont {S.}~\bibnamefont {Azzini}},
  \bibinfo {author} {\bibfnamefont {M.}~\bibnamefont {Liscidini}}, \bibinfo
  {author} {\bibfnamefont {M.}~\bibnamefont {Galli}}, \bibinfo {author}
  {\bibfnamefont {M.~J.}\ \bibnamefont {Strain}}, \bibinfo {author}
  {\bibfnamefont {M.}~\bibnamefont {Sorel}}, \bibinfo {author} {\bibfnamefont
  {J.~E.}\ \bibnamefont {Sipe}}, \ and\ \bibinfo {author} {\bibfnamefont
  {D.}~\bibnamefont {Bajoni}},\ }\href {\doibase 10.1364/optica.2.000088}
  {\bibfield  {journal} {\bibinfo  {journal} {Optica}\ }\textbf {\bibinfo
  {volume} {2}},\ \bibinfo {pages} {88} (\bibinfo {year} {2015})}\BibitemShut
  {NoStop}%
\bibitem [{\citenamefont {Hemsley}\ \emph {et~al.}(2016)\citenamefont
  {Hemsley}, \citenamefont {Bonneau}, \citenamefont {Pelc}, \citenamefont
  {Beausoleil}, \citenamefont {O'Brien},\ and\ \citenamefont
  {Thompson}}]{Hemsley2016}%
  \BibitemOpen
  \bibfield  {author} {\bibinfo {author} {\bibfnamefont {E.}~\bibnamefont
  {Hemsley}}, \bibinfo {author} {\bibfnamefont {D.}~\bibnamefont {Bonneau}},
  \bibinfo {author} {\bibfnamefont {J.}~\bibnamefont {Pelc}}, \bibinfo {author}
  {\bibfnamefont {R.}~\bibnamefont {Beausoleil}}, \bibinfo {author}
  {\bibfnamefont {J.~L.}\ \bibnamefont {O'Brien}}, \ and\ \bibinfo {author}
  {\bibfnamefont {M.~G.}\ \bibnamefont {Thompson}},\ }\href {\doibase
  10.1038/srep38908} {\bibfield  {journal} {\bibinfo  {journal} {Scientific
  reports}\ }\textbf {\bibinfo {volume} {6}},\ \bibinfo {pages} {38908}
  (\bibinfo {year} {2016})}\BibitemShut {NoStop}%
\bibitem [{\citenamefont {Ma}\ \emph {et~al.}(2017)\citenamefont {Ma},
  \citenamefont {Wang}, \citenamefont {Anant}, \citenamefont {Beyer},
  \citenamefont {Shaw},\ and\ \citenamefont {Mookherjea}}]{Ma2017}%
  \BibitemOpen
  \bibfield  {author} {\bibinfo {author} {\bibfnamefont {C.}~\bibnamefont
  {Ma}}, \bibinfo {author} {\bibfnamefont {X.}~\bibnamefont {Wang}}, \bibinfo
  {author} {\bibfnamefont {V.}~\bibnamefont {Anant}}, \bibinfo {author}
  {\bibfnamefont {A.~D.}\ \bibnamefont {Beyer}}, \bibinfo {author}
  {\bibfnamefont {M.~D.}\ \bibnamefont {Shaw}}, \ and\ \bibinfo {author}
  {\bibfnamefont {S.}~\bibnamefont {Mookherjea}},\ }\href {\doibase
  10.1364/oe.25.032995} {\bibfield  {journal} {\bibinfo  {journal} {Optics
  Express}\ }\textbf {\bibinfo {volume} {25}},\ \bibinfo {pages} {32995}
  (\bibinfo {year} {2017})},\ \Eprint {http://arxiv.org/abs/1710.01001}
  {1710.01001} \BibitemShut {NoStop}%
\bibitem [{\citenamefont {Samara}\ \emph {et~al.}(2019)\citenamefont {Samara},
  \citenamefont {Martin}, \citenamefont {Autebert}, \citenamefont {Karpov},
  \citenamefont {Kippenberg}, \citenamefont {Zbinden},\ and\ \citenamefont
  {Thew}}]{samara2019}%
  \BibitemOpen
  \bibfield  {author} {\bibinfo {author} {\bibfnamefont {F.}~\bibnamefont
  {Samara}}, \bibinfo {author} {\bibfnamefont {A.}~\bibnamefont {Martin}},
  \bibinfo {author} {\bibfnamefont {C.}~\bibnamefont {Autebert}}, \bibinfo
  {author} {\bibfnamefont {M.}~\bibnamefont {Karpov}}, \bibinfo {author}
  {\bibfnamefont {T.~J.}\ \bibnamefont {Kippenberg}}, \bibinfo {author}
  {\bibfnamefont {H.}~\bibnamefont {Zbinden}}, \ and\ \bibinfo {author}
  {\bibfnamefont {R.}~\bibnamefont {Thew}},\ }\href {\doibase
  10.1364/oe.27.019309} {\bibfield  {journal} {\bibinfo  {journal} {Optics
  Express}\ }\textbf {\bibinfo {volume} {27}},\ \bibinfo {pages} {19309}
  (\bibinfo {year} {2019})}\BibitemShut {NoStop}%
\bibitem [{\citenamefont {Oser}\ \emph {et~al.}(2020)\citenamefont {Oser},
  \citenamefont {Tanzilli}, \citenamefont {Mazeas}, \citenamefont
  {Alonso-Ramos}, \citenamefont {Roux}, \citenamefont {Sauder}, \citenamefont
  {Hua}, \citenamefont {Alibart}, \citenamefont {Vivien}, \citenamefont
  {Cassan},\ and\ \citenamefont {Labonté}}]{Oser2020}%
  \BibitemOpen
  \bibfield  {author} {\bibinfo {author} {\bibfnamefont {D.}~\bibnamefont
  {Oser}}, \bibinfo {author} {\bibfnamefont {S.}~\bibnamefont {Tanzilli}},
  \bibinfo {author} {\bibfnamefont {F.}~\bibnamefont {Mazeas}}, \bibinfo
  {author} {\bibfnamefont {C.}~\bibnamefont {Alonso-Ramos}}, \bibinfo {author}
  {\bibfnamefont {X.~L.}\ \bibnamefont {Roux}}, \bibinfo {author}
  {\bibfnamefont {G.}~\bibnamefont {Sauder}}, \bibinfo {author} {\bibfnamefont
  {X.}~\bibnamefont {Hua}}, \bibinfo {author} {\bibfnamefont {O.}~\bibnamefont
  {Alibart}}, \bibinfo {author} {\bibfnamefont {L.}~\bibnamefont {Vivien}},
  \bibinfo {author} {\bibfnamefont {E.}~\bibnamefont {Cassan}}, \ and\ \bibinfo
  {author} {\bibfnamefont {L.}~\bibnamefont {Labonté}},\ }\href {\doibase
  10.1038/s41534-020-0263-7} {\bibfield  {journal} {\bibinfo  {journal} {npj
  Quantum Information}\ }\textbf {\bibinfo {volume} {6}},\ \bibinfo {pages}
  {31} (\bibinfo {year} {2020})}\BibitemShut {NoStop}%
\bibitem [{\citenamefont {Pasquazi}\ \emph {et~al.}(2018)\citenamefont
  {Pasquazi}, \citenamefont {Peccianti}, \citenamefont {Razzari}, \citenamefont
  {Moss}, \citenamefont {Coen}, \citenamefont {Erkintalo}, \citenamefont
  {Chembo}, \citenamefont {Hansson}, \citenamefont {Wabnitz}, \citenamefont
  {Del'Haye}, \citenamefont {Xue}, \citenamefont {Weiner},\ and\ \citenamefont
  {Morandotti}}]{Pasquazi2018}%
  \BibitemOpen
  \bibfield  {author} {\bibinfo {author} {\bibfnamefont {A.}~\bibnamefont
  {Pasquazi}}, \bibinfo {author} {\bibfnamefont {M.}~\bibnamefont {Peccianti}},
  \bibinfo {author} {\bibfnamefont {L.}~\bibnamefont {Razzari}}, \bibinfo
  {author} {\bibfnamefont {D.~J.}\ \bibnamefont {Moss}}, \bibinfo {author}
  {\bibfnamefont {S.}~\bibnamefont {Coen}}, \bibinfo {author} {\bibfnamefont
  {M.}~\bibnamefont {Erkintalo}}, \bibinfo {author} {\bibfnamefont {Y.~K.}\
  \bibnamefont {Chembo}}, \bibinfo {author} {\bibfnamefont {T.}~\bibnamefont
  {Hansson}}, \bibinfo {author} {\bibfnamefont {S.}~\bibnamefont {Wabnitz}},
  \bibinfo {author} {\bibfnamefont {P.}~\bibnamefont {Del'Haye}}, \bibinfo
  {author} {\bibfnamefont {X.}~\bibnamefont {Xue}}, \bibinfo {author}
  {\bibfnamefont {A.~M.}\ \bibnamefont {Weiner}}, \ and\ \bibinfo {author}
  {\bibfnamefont {R.}~\bibnamefont {Morandotti}},\ }\href {\doibase
  10.1016/j.physrep.2017.08.004} {\bibfield  {journal} {\bibinfo  {journal}
  {Physics Reports}\ }\textbf {\bibinfo {volume} {729}},\ \bibinfo {pages} {1}
  (\bibinfo {year} {2018})}\BibitemShut {NoStop}%
\bibitem [{\citenamefont {Kues}\ \emph {et~al.}(2019)\citenamefont {Kues},
  \citenamefont {Reimer}, \citenamefont {Lukens}, \citenamefont {Munro},
  \citenamefont {Weiner}, \citenamefont {Moss},\ and\ \citenamefont
  {Morandotti}}]{Kues2019}%
  \BibitemOpen
  \bibfield  {author} {\bibinfo {author} {\bibfnamefont {M.}~\bibnamefont
  {Kues}}, \bibinfo {author} {\bibfnamefont {C.}~\bibnamefont {Reimer}},
  \bibinfo {author} {\bibfnamefont {J.~M.}\ \bibnamefont {Lukens}}, \bibinfo
  {author} {\bibfnamefont {W.~J.}\ \bibnamefont {Munro}}, \bibinfo {author}
  {\bibfnamefont {A.~M.}\ \bibnamefont {Weiner}}, \bibinfo {author}
  {\bibfnamefont {D.~J.}\ \bibnamefont {Moss}}, \ and\ \bibinfo {author}
  {\bibfnamefont {R.}~\bibnamefont {Morandotti}},\ }\href {\doibase
  10.1038/s41566-019-0363-0} {\bibfield  {journal} {\bibinfo  {journal} {Nature
  Photonics}\ }\textbf {\bibinfo {volume} {13}},\ \bibinfo {pages} {170}
  (\bibinfo {year} {2019})}\BibitemShut {NoStop}%
\bibitem [{\citenamefont {Llewellyn}\ \emph {et~al.}(2019)\citenamefont
  {Llewellyn}, \citenamefont {Ding}, \citenamefont {Faruque}, \citenamefont
  {Paesani}, \citenamefont {Bacco}, \citenamefont {Santagati}, \citenamefont
  {Qian}, \citenamefont {Li}, \citenamefont {Xiao}, \citenamefont {Huber},
  \citenamefont {Malik}, \citenamefont {Sinclair}, \citenamefont {Zhou},
  \citenamefont {Rottwitt}, \citenamefont {O’Brien}, \citenamefont {Rarity},
  \citenamefont {Gong}, \citenamefont {Oxenlowe}, \citenamefont {Wang},\ and\
  \citenamefont {Thompson}}]{Llewellyn2019}%
  \BibitemOpen
  \bibfield  {author} {\bibinfo {author} {\bibfnamefont {D.}~\bibnamefont
  {Llewellyn}}, \bibinfo {author} {\bibfnamefont {Y.}~\bibnamefont {Ding}},
  \bibinfo {author} {\bibfnamefont {I.~I.}\ \bibnamefont {Faruque}}, \bibinfo
  {author} {\bibfnamefont {S.}~\bibnamefont {Paesani}}, \bibinfo {author}
  {\bibfnamefont {D.}~\bibnamefont {Bacco}}, \bibinfo {author} {\bibfnamefont
  {R.}~\bibnamefont {Santagati}}, \bibinfo {author} {\bibfnamefont {Y.-J.}\
  \bibnamefont {Qian}}, \bibinfo {author} {\bibfnamefont {Y.}~\bibnamefont
  {Li}}, \bibinfo {author} {\bibfnamefont {Y.-F.}\ \bibnamefont {Xiao}},
  \bibinfo {author} {\bibfnamefont {M.}~\bibnamefont {Huber}}, \bibinfo
  {author} {\bibfnamefont {M.}~\bibnamefont {Malik}}, \bibinfo {author}
  {\bibfnamefont {G.~F.}\ \bibnamefont {Sinclair}}, \bibinfo {author}
  {\bibfnamefont {X.}~\bibnamefont {Zhou}}, \bibinfo {author} {\bibfnamefont
  {K.}~\bibnamefont {Rottwitt}}, \bibinfo {author} {\bibfnamefont {J.~L.}\
  \bibnamefont {O’Brien}}, \bibinfo {author} {\bibfnamefont {J.~G.}\
  \bibnamefont {Rarity}}, \bibinfo {author} {\bibfnamefont {Q.}~\bibnamefont
  {Gong}}, \bibinfo {author} {\bibfnamefont {L.~K.}\ \bibnamefont {Oxenlowe}},
  \bibinfo {author} {\bibfnamefont {J.}~\bibnamefont {Wang}}, \ and\ \bibinfo
  {author} {\bibfnamefont {M.~G.}\ \bibnamefont {Thompson}},\ }\href {\doibase
  10.1038/s41567-019-0727-x} {\bibfield  {journal} {\bibinfo  {journal} {Nature
  Physics}\ }\textbf {\bibinfo {volume} {16}},\ \bibinfo {pages} {148}
  (\bibinfo {year} {2019})},\ \Eprint {http://arxiv.org/abs/1911.07839}
  {1911.07839} \BibitemShut {NoStop}%
\bibitem [{\citenamefont {Aboussouan}\ \emph {et~al.}(2010)\citenamefont
  {Aboussouan}, \citenamefont {Alibart}, \citenamefont {Ostrowsky},
  \citenamefont {Baldi},\ and\ \citenamefont {Tanzilli}}]{Tanzilli2010}%
  \BibitemOpen
  \bibfield  {author} {\bibinfo {author} {\bibfnamefont {P.}~\bibnamefont
  {Aboussouan}}, \bibinfo {author} {\bibfnamefont {O.}~\bibnamefont {Alibart}},
  \bibinfo {author} {\bibfnamefont {D.~B.}\ \bibnamefont {Ostrowsky}}, \bibinfo
  {author} {\bibfnamefont {P.}~\bibnamefont {Baldi}}, \ and\ \bibinfo {author}
  {\bibfnamefont {S.}~\bibnamefont {Tanzilli}},\ }\href {\doibase
  10.1103/PhysRevA.81.021801} {\bibfield  {journal} {\bibinfo  {journal}
  {Physical Review A}\ }\textbf {\bibinfo {volume} {81}},\ \bibinfo {pages}
  {021801} (\bibinfo {year} {2010})}\BibitemShut {NoStop}%
\bibitem [{\citenamefont {Kaltenbaek}\ \emph {et~al.}(2006)\citenamefont
  {Kaltenbaek}, \citenamefont {Blauensteiner}, \citenamefont
  {\ifmmode~\dot{Z}\else \.{Z}\fi{}ukowski}, \citenamefont {Aspelmeyer},\ and\
  \citenamefont {Zeilinger}}]{Kaltenbaek2006}%
  \BibitemOpen
  \bibfield  {author} {\bibinfo {author} {\bibfnamefont {R.}~\bibnamefont
  {Kaltenbaek}}, \bibinfo {author} {\bibfnamefont {B.}~\bibnamefont
  {Blauensteiner}}, \bibinfo {author} {\bibfnamefont {M.}~\bibnamefont
  {\ifmmode~\dot{Z}\else \.{Z}\fi{}ukowski}}, \bibinfo {author} {\bibfnamefont
  {M.}~\bibnamefont {Aspelmeyer}}, \ and\ \bibinfo {author} {\bibfnamefont
  {A.}~\bibnamefont {Zeilinger}},\ }\href {\doibase
  10.1103/physrevlett.96.240502} {\bibfield  {journal} {\bibinfo  {journal}
  {Physical Review Letters}\ }\textbf {\bibinfo {volume} {96}},\ \bibinfo
  {pages} {240502} (\bibinfo {year} {2006})}\BibitemShut {NoStop}%
\bibitem [{\citenamefont {Yang}\ \emph {et~al.}(2006)\citenamefont {Yang},
  \citenamefont {Zhang}, \citenamefont {Chen}, \citenamefont {Lu},
  \citenamefont {Yin}, \citenamefont {Pan}, \citenamefont {Wei}, \citenamefont
  {Tian},\ and\ \citenamefont {Zhang}}]{Yang2006}%
  \BibitemOpen
  \bibfield  {author} {\bibinfo {author} {\bibfnamefont {T.}~\bibnamefont
  {Yang}}, \bibinfo {author} {\bibfnamefont {Q.}~\bibnamefont {Zhang}},
  \bibinfo {author} {\bibfnamefont {T.-Y.}\ \bibnamefont {Chen}}, \bibinfo
  {author} {\bibfnamefont {S.}~\bibnamefont {Lu}}, \bibinfo {author}
  {\bibfnamefont {J.}~\bibnamefont {Yin}}, \bibinfo {author} {\bibfnamefont
  {J.-W.}\ \bibnamefont {Pan}}, \bibinfo {author} {\bibfnamefont {Z.-Y.}\
  \bibnamefont {Wei}}, \bibinfo {author} {\bibfnamefont {J.-R.}\ \bibnamefont
  {Tian}}, \ and\ \bibinfo {author} {\bibfnamefont {J.}~\bibnamefont {Zhang}},\
  }\href {\doibase 10.1103/physrevlett.96.110501} {\bibfield  {journal}
  {\bibinfo  {journal} {Physical Review Letters}\ }\textbf {\bibinfo {volume}
  {96}},\ \bibinfo {pages} {110501} (\bibinfo {year} {2006})},\ \Eprint
  {http://arxiv.org/abs/quant-ph/0502146} {quant-ph/0502146} \BibitemShut
  {NoStop}%
\bibitem [{\citenamefont {Kaltenbaek}\ \emph {et~al.}(2009)\citenamefont
  {Kaltenbaek}, \citenamefont {Prevedel}, \citenamefont {Aspelmeyer},\ and\
  \citenamefont {Zeilinger}}]{Kaltenbaek2009}%
  \BibitemOpen
  \bibfield  {author} {\bibinfo {author} {\bibfnamefont {R.}~\bibnamefont
  {Kaltenbaek}}, \bibinfo {author} {\bibfnamefont {R.}~\bibnamefont
  {Prevedel}}, \bibinfo {author} {\bibfnamefont {M.}~\bibnamefont
  {Aspelmeyer}}, \ and\ \bibinfo {author} {\bibfnamefont {A.}~\bibnamefont
  {Zeilinger}},\ }\href {\doibase 10.1103/physreva.79.040302} {\bibfield
  {journal} {\bibinfo  {journal} {Physical Review A}\ }\textbf {\bibinfo
  {volume} {79}},\ \bibinfo {pages} {040302} (\bibinfo {year}
  {2009})}\BibitemShut {NoStop}%
\bibitem [{\citenamefont {Sun}\ \emph {et~al.}(2017{\natexlab{a}})\citenamefont
  {Sun}, \citenamefont {Mao}, \citenamefont {Jiang}, \citenamefont {Zhao},
  \citenamefont {Chen}, \citenamefont {Zhang}, \citenamefont {Zhang},
  \citenamefont {Jiang}, \citenamefont {Chen}, \citenamefont {You},
  \citenamefont {Li}, \citenamefont {Huang}, \citenamefont {Chen},
  \citenamefont {Wang}, \citenamefont {Ma}, \citenamefont {Zhang},\ and\
  \citenamefont {Pan}}]{sun2017}%
  \BibitemOpen
  \bibfield  {author} {\bibinfo {author} {\bibfnamefont {Q.-C.}\ \bibnamefont
  {Sun}}, \bibinfo {author} {\bibfnamefont {Y.-L.}\ \bibnamefont {Mao}},
  \bibinfo {author} {\bibfnamefont {Y.-F.}\ \bibnamefont {Jiang}}, \bibinfo
  {author} {\bibfnamefont {Q.}~\bibnamefont {Zhao}}, \bibinfo {author}
  {\bibfnamefont {S.-J.}\ \bibnamefont {Chen}}, \bibinfo {author}
  {\bibfnamefont {W.}~\bibnamefont {Zhang}}, \bibinfo {author} {\bibfnamefont
  {W.-J.}\ \bibnamefont {Zhang}}, \bibinfo {author} {\bibfnamefont
  {X.}~\bibnamefont {Jiang}}, \bibinfo {author} {\bibfnamefont {T.-Y.}\
  \bibnamefont {Chen}}, \bibinfo {author} {\bibfnamefont {L.-X.}\ \bibnamefont
  {You}}, \bibinfo {author} {\bibfnamefont {L.}~\bibnamefont {Li}}, \bibinfo
  {author} {\bibfnamefont {Y.-D.}\ \bibnamefont {Huang}}, \bibinfo {author}
  {\bibfnamefont {X.-F.}\ \bibnamefont {Chen}}, \bibinfo {author}
  {\bibfnamefont {Z.}~\bibnamefont {Wang}}, \bibinfo {author} {\bibfnamefont
  {X.}~\bibnamefont {Ma}}, \bibinfo {author} {\bibfnamefont {Q.}~\bibnamefont
  {Zhang}}, \ and\ \bibinfo {author} {\bibfnamefont {J.-W.}\ \bibnamefont
  {Pan}},\ }\href {\doibase 10.1103/physreva.95.032306} {\bibfield  {journal}
  {\bibinfo  {journal} {Physical Review A}\ }\textbf {\bibinfo {volume} {95}},\
  \bibinfo {pages} {032306} (\bibinfo {year} {2017}{\natexlab{a}})}\BibitemShut
  {NoStop}%
\bibitem [{\citenamefont {Sun}\ \emph {et~al.}(2017{\natexlab{b}})\citenamefont
  {Sun}, \citenamefont {Jiang}, \citenamefont {Mao}, \citenamefont {You},
  \citenamefont {Zhang}, \citenamefont {Zhang}, \citenamefont {Jiang},
  \citenamefont {Chen}, \citenamefont {Li}, \citenamefont {Huang},
  \citenamefont {Chen}, \citenamefont {Wang}, \citenamefont {Fan},
  \citenamefont {Zhang},\ and\ \citenamefont {Pan}}]{Sun2017_100km}%
  \BibitemOpen
  \bibfield  {author} {\bibinfo {author} {\bibfnamefont {Q.-C.}\ \bibnamefont
  {Sun}}, \bibinfo {author} {\bibfnamefont {Y.-F.}\ \bibnamefont {Jiang}},
  \bibinfo {author} {\bibfnamefont {Y.-L.}\ \bibnamefont {Mao}}, \bibinfo
  {author} {\bibfnamefont {L.-X.}\ \bibnamefont {You}}, \bibinfo {author}
  {\bibfnamefont {W.}~\bibnamefont {Zhang}}, \bibinfo {author} {\bibfnamefont
  {W.-J.}\ \bibnamefont {Zhang}}, \bibinfo {author} {\bibfnamefont
  {X.}~\bibnamefont {Jiang}}, \bibinfo {author} {\bibfnamefont {T.-Y.}\
  \bibnamefont {Chen}}, \bibinfo {author} {\bibfnamefont {H.}~\bibnamefont
  {Li}}, \bibinfo {author} {\bibfnamefont {Y.-D.}\ \bibnamefont {Huang}},
  \bibinfo {author} {\bibfnamefont {X.-F.}\ \bibnamefont {Chen}}, \bibinfo
  {author} {\bibfnamefont {Z.}~\bibnamefont {Wang}}, \bibinfo {author}
  {\bibfnamefont {J.}~\bibnamefont {Fan}}, \bibinfo {author} {\bibfnamefont
  {Q.}~\bibnamefont {Zhang}}, \ and\ \bibinfo {author} {\bibfnamefont {J.-W.}\
  \bibnamefont {Pan}},\ }\href {\doibase 10.1364/optica.4.001214} {\bibfield
  {journal} {\bibinfo  {journal} {Optica}\ }\textbf {\bibinfo {volume} {4}},\
  \bibinfo {pages} {1214} (\bibinfo {year} {2017}{\natexlab{b}})}\BibitemShut
  {NoStop}%
\bibitem [{\citenamefont {Halder}\ \emph {et~al.}(2007)\citenamefont {Halder},
  \citenamefont {Beveratos}, \citenamefont {Gisin}, \citenamefont {Scarani},
  \citenamefont {Simon},\ and\ \citenamefont {Zbinden}}]{Halder2007}%
  \BibitemOpen
  \bibfield  {author} {\bibinfo {author} {\bibfnamefont {M.}~\bibnamefont
  {Halder}}, \bibinfo {author} {\bibfnamefont {A.}~\bibnamefont {Beveratos}},
  \bibinfo {author} {\bibfnamefont {N.}~\bibnamefont {Gisin}}, \bibinfo
  {author} {\bibfnamefont {V.}~\bibnamefont {Scarani}}, \bibinfo {author}
  {\bibfnamefont {C.}~\bibnamefont {Simon}}, \ and\ \bibinfo {author}
  {\bibfnamefont {H.}~\bibnamefont {Zbinden}},\ }\href {\doibase
  10.1038/nphys700} {\bibfield  {journal} {\bibinfo  {journal} {Nature
  Physics}\ }\textbf {\bibinfo {volume} {3}},\ \bibinfo {pages} {692} (\bibinfo
  {year} {2007})}\BibitemShut {NoStop}%
\bibitem [{\citenamefont {Osorio}\ \emph {et~al.}(2013)\citenamefont {Osorio},
  \citenamefont {Sangouard},\ and\ \citenamefont {Thew}}]{Osorio2013}%
  \BibitemOpen
  \bibfield  {author} {\bibinfo {author} {\bibfnamefont {C.~I.}\ \bibnamefont
  {Osorio}}, \bibinfo {author} {\bibfnamefont {N.}~\bibnamefont {Sangouard}}, \
  and\ \bibinfo {author} {\bibfnamefont {R.~T.}\ \bibnamefont {Thew}},\ }\href
  {\doibase 10.1088/0953-4075/46/5/055501} {\bibfield  {journal} {\bibinfo
  {journal} {Journal of Physics B: Atomic, Molecular and Optical Physics}\
  }\textbf {\bibinfo {volume} {46}},\ \bibinfo {pages} {055501} (\bibinfo
  {year} {2013})},\ \Eprint {http://arxiv.org/abs/1211.0120} {1211.0120}
  \BibitemShut {NoStop}%
\bibitem [{\citenamefont {Liu}\ \emph {et~al.}(2020)\citenamefont {Liu},
  \citenamefont {Huang}, \citenamefont {Wang}, \citenamefont {He},
  \citenamefont {Raja}, \citenamefont {Liu}, \citenamefont {Engelsen},\ and\
  \citenamefont {Kippenberg}}]{liu2020high}%
  \BibitemOpen
  \bibfield  {author} {\bibinfo {author} {\bibfnamefont {J.}~\bibnamefont
  {Liu}}, \bibinfo {author} {\bibfnamefont {G.}~\bibnamefont {Huang}}, \bibinfo
  {author} {\bibfnamefont {R.~N.}\ \bibnamefont {Wang}}, \bibinfo {author}
  {\bibfnamefont {J.}~\bibnamefont {He}}, \bibinfo {author} {\bibfnamefont
  {A.~S.}\ \bibnamefont {Raja}}, \bibinfo {author} {\bibfnamefont
  {T.}~\bibnamefont {Liu}}, \bibinfo {author} {\bibfnamefont {N.~J.}\
  \bibnamefont {Engelsen}}, \ and\ \bibinfo {author} {\bibfnamefont {T.~J.}\
  \bibnamefont {Kippenberg}},\ }\href@noop {} {\bibfield  {journal} {\bibinfo
  {journal} {arXiv preprint arXiv:2005.13949}\ } (\bibinfo {year}
  {2020})}\BibitemShut {NoStop}%
\bibitem [{\citenamefont {Gyger}\ \emph {et~al.}(2020)\citenamefont {Gyger},
  \citenamefont {Liu}, \citenamefont {Yang}, \citenamefont {He}, \citenamefont
  {Raja}, \citenamefont {Wang}, \citenamefont {Bhave}, \citenamefont
  {Kippenberg},\ and\ \citenamefont {Th{\'e}venaz}}]{gyger2020observation}%
  \BibitemOpen
  \bibfield  {author} {\bibinfo {author} {\bibfnamefont {F.}~\bibnamefont
  {Gyger}}, \bibinfo {author} {\bibfnamefont {J.}~\bibnamefont {Liu}}, \bibinfo
  {author} {\bibfnamefont {F.}~\bibnamefont {Yang}}, \bibinfo {author}
  {\bibfnamefont {J.}~\bibnamefont {He}}, \bibinfo {author} {\bibfnamefont
  {A.~S.}\ \bibnamefont {Raja}}, \bibinfo {author} {\bibfnamefont {R.~N.}\
  \bibnamefont {Wang}}, \bibinfo {author} {\bibfnamefont {S.~A.}\ \bibnamefont
  {Bhave}}, \bibinfo {author} {\bibfnamefont {T.~J.}\ \bibnamefont
  {Kippenberg}}, \ and\ \bibinfo {author} {\bibfnamefont {L.}~\bibnamefont
  {Th{\'e}venaz}},\ }\href@noop {} {\bibfield  {journal} {\bibinfo  {journal}
  {Physical Review Letters}\ }\textbf {\bibinfo {volume} {124}},\ \bibinfo
  {pages} {013902} (\bibinfo {year} {2020})}\BibitemShut {NoStop}%
\bibitem [{\citenamefont {Brasch}\ \emph {et~al.}(2014)\citenamefont {Brasch},
  \citenamefont {Chen}, \citenamefont {Schiller},\ and\ \citenamefont
  {Kippenberg}}]{Brasch:14}%
  \BibitemOpen
  \bibfield  {author} {\bibinfo {author} {\bibfnamefont {V.}~\bibnamefont
  {Brasch}}, \bibinfo {author} {\bibfnamefont {Q.-F.}\ \bibnamefont {Chen}},
  \bibinfo {author} {\bibfnamefont {S.}~\bibnamefont {Schiller}}, \ and\
  \bibinfo {author} {\bibfnamefont {T.~J.}\ \bibnamefont {Kippenberg}},\ }\href
  {\doibase 10.1364/OE.22.030786} {\bibfield  {journal} {\bibinfo  {journal}
  {Opt. Express}\ }\textbf {\bibinfo {volume} {22}},\ \bibinfo {pages} {30786}
  (\bibinfo {year} {2014})}\BibitemShut {NoStop}%
\bibitem [{\citenamefont {Kippenberg}\ \emph {et~al.}(2018)\citenamefont
  {Kippenberg}, \citenamefont {Gaeta}, \citenamefont {Lipson},\ and\
  \citenamefont {Gorodetsky}}]{kippenberg2018}%
  \BibitemOpen
  \bibfield  {author} {\bibinfo {author} {\bibfnamefont {T.~J.}\ \bibnamefont
  {Kippenberg}}, \bibinfo {author} {\bibfnamefont {A.~L.}\ \bibnamefont
  {Gaeta}}, \bibinfo {author} {\bibfnamefont {M.}~\bibnamefont {Lipson}}, \
  and\ \bibinfo {author} {\bibfnamefont {M.~L.}\ \bibnamefont {Gorodetsky}},\
  }\href {http://science.sciencemag.org/content/361/6402/eaan8083} {\bibfield
  {journal} {\bibinfo  {journal} {Science}\ }\textbf {\bibinfo {volume} {361}}
  (\bibinfo {year} {2018})}\BibitemShut {NoStop}%
\bibitem [{\citenamefont {Pfeiffer}\ \emph {et~al.}(2016)\citenamefont
  {Pfeiffer}, \citenamefont {Kordts}, \citenamefont {Brasch}, \citenamefont
  {Zervas}, \citenamefont {Geiselmann}, \citenamefont {Jost},\ and\
  \citenamefont {Kippenberg}}]{Pfeiffer2016}%
  \BibitemOpen
  \bibfield  {author} {\bibinfo {author} {\bibfnamefont {M.~H.~P.}\
  \bibnamefont {Pfeiffer}}, \bibinfo {author} {\bibfnamefont {A.}~\bibnamefont
  {Kordts}}, \bibinfo {author} {\bibfnamefont {V.}~\bibnamefont {Brasch}},
  \bibinfo {author} {\bibfnamefont {M.}~\bibnamefont {Zervas}}, \bibinfo
  {author} {\bibfnamefont {M.}~\bibnamefont {Geiselmann}}, \bibinfo {author}
  {\bibfnamefont {J.~D.}\ \bibnamefont {Jost}}, \ and\ \bibinfo {author}
  {\bibfnamefont {T.~J.}\ \bibnamefont {Kippenberg}},\ }\href {\doibase
  10.1364/optica.3.000020} {\bibfield  {journal} {\bibinfo  {journal} {Optica}\
  }\textbf {\bibinfo {volume} {3}},\ \bibinfo {pages} {20} (\bibinfo {year}
  {2016})},\ \Eprint {http://arxiv.org/abs/1511.05716} {1511.05716}
  \BibitemShut {NoStop}%
\bibitem [{\citenamefont {Hong}\ \emph {et~al.}(1987)\citenamefont {Hong},
  \citenamefont {Ou},\ and\ \citenamefont {Mandel}}]{Hong1987}%
  \BibitemOpen
  \bibfield  {author} {\bibinfo {author} {\bibfnamefont {C.~K.}\ \bibnamefont
  {Hong}}, \bibinfo {author} {\bibfnamefont {Z.~Y.}\ \bibnamefont {Ou}}, \ and\
  \bibinfo {author} {\bibfnamefont {L.}~\bibnamefont {Mandel}},\ }\href
  {\doibase 10.1103/physrevlett.59.2044} {\bibfield  {journal} {\bibinfo
  {journal} {Physical Review Letters}\ }\textbf {\bibinfo {volume} {59}},\
  \bibinfo {pages} {2044} (\bibinfo {year} {1987})}\BibitemShut {NoStop}%
\bibitem [{\citenamefont {Marcikic}\ \emph {et~al.}(2004)\citenamefont
  {Marcikic}, \citenamefont {de~Riedmatten}, \citenamefont {Tittel},
  \citenamefont {Zbinden}, \citenamefont {Legr\'e},\ and\ \citenamefont
  {Gisin}}]{Marcikic2004}%
  \BibitemOpen
  \bibfield  {author} {\bibinfo {author} {\bibfnamefont {I.}~\bibnamefont
  {Marcikic}}, \bibinfo {author} {\bibfnamefont {H.}~\bibnamefont
  {de~Riedmatten}}, \bibinfo {author} {\bibfnamefont {W.}~\bibnamefont
  {Tittel}}, \bibinfo {author} {\bibfnamefont {H.}~\bibnamefont {Zbinden}},
  \bibinfo {author} {\bibfnamefont {M.}~\bibnamefont {Legr\'e}}, \ and\
  \bibinfo {author} {\bibfnamefont {N.}~\bibnamefont {Gisin}},\ }\href
  {\doibase 10.1103/PhysRevLett.93.180502} {\bibfield  {journal} {\bibinfo
  {journal} {Phys. Rev. Lett.}\ }\textbf {\bibinfo {volume} {93}},\ \bibinfo
  {pages} {180502} (\bibinfo {year} {2004})}\BibitemShut {NoStop}%
\bibitem [{\citenamefont {Spring}\ \emph {et~al.}(2017)\citenamefont {Spring},
  \citenamefont {Mennea}, \citenamefont {Metcalf}, \citenamefont {Humphreys},
  \citenamefont {Gates}, \citenamefont {Rogers}, \citenamefont {S\"{o}ller},
  \citenamefont {Smith}, \citenamefont {Kolthammer}, \citenamefont {Smith},\
  and\ \citenamefont {Walmsley}}]{Spring2017}%
  \BibitemOpen
  \bibfield  {author} {\bibinfo {author} {\bibfnamefont {J.~B.}\ \bibnamefont
  {Spring}}, \bibinfo {author} {\bibfnamefont {P.~L.}\ \bibnamefont {Mennea}},
  \bibinfo {author} {\bibfnamefont {B.~J.}\ \bibnamefont {Metcalf}}, \bibinfo
  {author} {\bibfnamefont {P.~C.}\ \bibnamefont {Humphreys}}, \bibinfo {author}
  {\bibfnamefont {J.~C.}\ \bibnamefont {Gates}}, \bibinfo {author}
  {\bibfnamefont {H.~L.}\ \bibnamefont {Rogers}}, \bibinfo {author}
  {\bibfnamefont {C.}~\bibnamefont {S\"{o}ller}}, \bibinfo {author}
  {\bibfnamefont {B.~J.}\ \bibnamefont {Smith}}, \bibinfo {author}
  {\bibfnamefont {W.~S.}\ \bibnamefont {Kolthammer}}, \bibinfo {author}
  {\bibfnamefont {P.~G.~R.}\ \bibnamefont {Smith}}, \ and\ \bibinfo {author}
  {\bibfnamefont {I.~A.}\ \bibnamefont {Walmsley}},\ }\href {\doibase
  10.1364/optica.4.000090} {\bibfield  {journal} {\bibinfo  {journal} {Optica}\
  }\textbf {\bibinfo {volume} {4}},\ \bibinfo {pages} {90} (\bibinfo {year}
  {2017})},\ \Eprint {http://arxiv.org/abs/1603.06984} {1603.06984}
  \BibitemShut {NoStop}%
\bibitem [{\citenamefont {Zhang}\ \emph {et~al.}(2016)\citenamefont {Zhang},
  \citenamefont {Jiang}, \citenamefont {Bell}, \citenamefont {Choi},
  \citenamefont {Chae},\ and\ \citenamefont {Xiong}}]{Zhang2016}%
  \BibitemOpen
  \bibfield  {author} {\bibinfo {author} {\bibfnamefont {X.}~\bibnamefont
  {Zhang}}, \bibinfo {author} {\bibfnamefont {R.}~\bibnamefont {Jiang}},
  \bibinfo {author} {\bibfnamefont {B.}~\bibnamefont {Bell}}, \bibinfo {author}
  {\bibfnamefont {D.-Y.}\ \bibnamefont {Choi}}, \bibinfo {author}
  {\bibfnamefont {C.}~\bibnamefont {Chae}}, \ and\ \bibinfo {author}
  {\bibfnamefont {C.}~\bibnamefont {Xiong}},\ }\href {\doibase
  10.3390/technologies4030025} {\bibfield  {journal} {\bibinfo  {journal}
  {Technologies}\ }\textbf {\bibinfo {volume} {4}},\ \bibinfo {pages} {25}
  (\bibinfo {year} {2016})}\BibitemShut {NoStop}%
\bibitem [{\citenamefont {Harada}\ \emph {et~al.}(2011)\citenamefont {Harada},
  \citenamefont {Takesue}, \citenamefont {Fukuda}, \citenamefont {Tsuchizawa},
  \citenamefont {Watanabe}, \citenamefont {Yamada}, \citenamefont {Tokura},\
  and\ \citenamefont {Itabashi}}]{Harada2011}%
  \BibitemOpen
  \bibfield  {author} {\bibinfo {author} {\bibfnamefont {K.-i.}\ \bibnamefont
  {Harada}}, \bibinfo {author} {\bibfnamefont {H.}~\bibnamefont {Takesue}},
  \bibinfo {author} {\bibfnamefont {H.}~\bibnamefont {Fukuda}}, \bibinfo
  {author} {\bibfnamefont {T.}~\bibnamefont {Tsuchizawa}}, \bibinfo {author}
  {\bibfnamefont {T.}~\bibnamefont {Watanabe}}, \bibinfo {author}
  {\bibfnamefont {K.}~\bibnamefont {Yamada}}, \bibinfo {author} {\bibfnamefont
  {Y.}~\bibnamefont {Tokura}}, \ and\ \bibinfo {author} {\bibfnamefont {S.-i.}\
  \bibnamefont {Itabashi}},\ }\href {\doibase 10.1088/1367-2630/13/6/065005}
  {\bibfield  {journal} {\bibinfo  {journal} {New Journal of Physics}\ }\textbf
  {\bibinfo {volume} {13}},\ \bibinfo {pages} {065005} (\bibinfo {year}
  {2011})}\BibitemShut {NoStop}%
\bibitem [{\citenamefont {Sekatski}\ \emph {et~al.}(2012)\citenamefont
  {Sekatski}, \citenamefont {Sangouard}, \citenamefont {Bussi{\`{e}}res},
  \citenamefont {Clausen}, \citenamefont {Gisin},\ and\ \citenamefont
  {Zbinden}}]{Sekatski2012}%
  \BibitemOpen
  \bibfield  {author} {\bibinfo {author} {\bibfnamefont {P.}~\bibnamefont
  {Sekatski}}, \bibinfo {author} {\bibfnamefont {N.}~\bibnamefont {Sangouard}},
  \bibinfo {author} {\bibfnamefont {F.}~\bibnamefont {Bussi{\`{e}}res}},
  \bibinfo {author} {\bibfnamefont {C.}~\bibnamefont {Clausen}}, \bibinfo
  {author} {\bibfnamefont {N.}~\bibnamefont {Gisin}}, \ and\ \bibinfo {author}
  {\bibfnamefont {H.}~\bibnamefont {Zbinden}},\ }\href {\doibase
  10.1088/0953-4075/45/12/124016} {\bibfield  {journal} {\bibinfo  {journal}
  {Journal of Physics B: Atomic, Molecular and Optical Physics}\ }\textbf
  {\bibinfo {volume} {45}},\ \bibinfo {pages} {124016} (\bibinfo {year}
  {2012})},\ \Eprint {http://arxiv.org/abs/1109.0194} {1109.0194} \BibitemShut
  {NoStop}%
\bibitem [{\citenamefont {Helt}\ \emph {et~al.}(2010)\citenamefont {Helt},
  \citenamefont {Yang}, \citenamefont {Liscidini},\ and\ \citenamefont
  {Sipe}}]{Helt2010}%
  \BibitemOpen
  \bibfield  {author} {\bibinfo {author} {\bibfnamefont {L.~G.}\ \bibnamefont
  {Helt}}, \bibinfo {author} {\bibfnamefont {Z.}~\bibnamefont {Yang}}, \bibinfo
  {author} {\bibfnamefont {M.}~\bibnamefont {Liscidini}}, \ and\ \bibinfo
  {author} {\bibfnamefont {J.~E.}\ \bibnamefont {Sipe}},\ }\href {\doibase
  10.1364/ol.35.003006} {\bibfield  {journal} {\bibinfo  {journal} {Optics
  letters}\ }\textbf {\bibinfo {volume} {35}},\ \bibinfo {pages} {3006}
  (\bibinfo {year} {2010})}\BibitemShut {NoStop}%
\bibitem [{\citenamefont {Huang}\ \emph {et~al.}(2010)\citenamefont {Huang},
  \citenamefont {Altepeter},\ and\ \citenamefont {Kumar}}]{Huang2010}%
  \BibitemOpen
  \bibfield  {author} {\bibinfo {author} {\bibfnamefont {Y.-P.}\ \bibnamefont
  {Huang}}, \bibinfo {author} {\bibfnamefont {J.~B.}\ \bibnamefont
  {Altepeter}}, \ and\ \bibinfo {author} {\bibfnamefont {P.}~\bibnamefont
  {Kumar}},\ }\href {\doibase 10.1103/physreva.82.043826} {\bibfield  {journal}
  {\bibinfo  {journal} {Physical Review A}\ }\textbf {\bibinfo {volume} {82}},\
  \bibinfo {pages} {043826} (\bibinfo {year} {2010})},\ \Eprint
  {http://arxiv.org/abs/1008.2792} {1008.2792} \BibitemShut {NoStop}%
\bibitem [{\citenamefont {Tison}\ \emph {et~al.}(2017)\citenamefont {Tison},
  \citenamefont {Steidle}, \citenamefont {Fanto}, \citenamefont {Wang},
  \citenamefont {Mogent}, \citenamefont {Rizzo}, \citenamefont {Preble},\ and\
  \citenamefont {Alsing}}]{Tison2017}%
  \BibitemOpen
  \bibfield  {author} {\bibinfo {author} {\bibfnamefont {C.~C.}\ \bibnamefont
  {Tison}}, \bibinfo {author} {\bibfnamefont {J.~A.}\ \bibnamefont {Steidle}},
  \bibinfo {author} {\bibfnamefont {M.~L.}\ \bibnamefont {Fanto}}, \bibinfo
  {author} {\bibfnamefont {Z.}~\bibnamefont {Wang}}, \bibinfo {author}
  {\bibfnamefont {N.~A.}\ \bibnamefont {Mogent}}, \bibinfo {author}
  {\bibfnamefont {A.}~\bibnamefont {Rizzo}}, \bibinfo {author} {\bibfnamefont
  {S.~F.}\ \bibnamefont {Preble}}, \ and\ \bibinfo {author} {\bibfnamefont
  {P.~M.}\ \bibnamefont {Alsing}},\ }\href {\doibase 10.1364/oe.25.033088}
  {\bibfield  {journal} {\bibinfo  {journal} {Optics Express}\ }\textbf
  {\bibinfo {volume} {25}},\ \bibinfo {pages} {33088} (\bibinfo {year}
  {2017})},\ \Eprint {http://arxiv.org/abs/1703.08368} {1703.08368}
  \BibitemShut {NoStop}%
\bibitem [{\citenamefont {Vernon}\ \emph {et~al.}(2016)\citenamefont {Vernon},
  \citenamefont {Liscidini},\ and\ \citenamefont {Sipe}}]{Vernon2016}%
  \BibitemOpen
  \bibfield  {author} {\bibinfo {author} {\bibfnamefont {Z.}~\bibnamefont
  {Vernon}}, \bibinfo {author} {\bibfnamefont {M.}~\bibnamefont {Liscidini}}, \
  and\ \bibinfo {author} {\bibfnamefont {J.~E.}\ \bibnamefont {Sipe}},\ }\href
  {\doibase 10.1364/OL.41.000788} {\bibfield  {journal} {\bibinfo  {journal}
  {Opt. Lett.}\ }\textbf {\bibinfo {volume} {41}},\ \bibinfo {pages} {788}
  (\bibinfo {year} {2016})}\BibitemShut {NoStop}%
\bibitem [{\citenamefont {Liu}\ \emph {et~al.}(2018)\citenamefont {Liu},
  \citenamefont {Raja}, \citenamefont {Pfeiffer}, \citenamefont {Herkommer},
  \citenamefont {Guo}, \citenamefont {Zervas}, \citenamefont {Geiselmann},\
  and\ \citenamefont {Kippenberg}}]{Liu18}%
  \BibitemOpen
  \bibfield  {author} {\bibinfo {author} {\bibfnamefont {J.}~\bibnamefont
  {Liu}}, \bibinfo {author} {\bibfnamefont {A.~S.}\ \bibnamefont {Raja}},
  \bibinfo {author} {\bibfnamefont {M.~H.~P.}\ \bibnamefont {Pfeiffer}},
  \bibinfo {author} {\bibfnamefont {C.}~\bibnamefont {Herkommer}}, \bibinfo
  {author} {\bibfnamefont {H.}~\bibnamefont {Guo}}, \bibinfo {author}
  {\bibfnamefont {M.}~\bibnamefont {Zervas}}, \bibinfo {author} {\bibfnamefont
  {M.}~\bibnamefont {Geiselmann}}, \ and\ \bibinfo {author} {\bibfnamefont
  {T.~J.}\ \bibnamefont {Kippenberg}},\ }\href {\doibase 10.1364/OL.43.003200}
  {\bibfield  {journal} {\bibinfo  {journal} {Opt. Lett.}\ }\textbf {\bibinfo
  {volume} {43}},\ \bibinfo {pages} {3200} (\bibinfo {year}
  {2018})}\BibitemShut {NoStop}%
\bibitem [{\citenamefont {Bruno}\ \emph {et~al.}(2014)\citenamefont {Bruno},
  \citenamefont {Martin}, \citenamefont {Guerreiro}, \citenamefont
  {Sanguinetti},\ and\ \citenamefont {Thew}}]{Bruno2014}%
  \BibitemOpen
  \bibfield  {author} {\bibinfo {author} {\bibfnamefont {N.}~\bibnamefont
  {Bruno}}, \bibinfo {author} {\bibfnamefont {A.}~\bibnamefont {Martin}},
  \bibinfo {author} {\bibfnamefont {T.}~\bibnamefont {Guerreiro}}, \bibinfo
  {author} {\bibfnamefont {B.}~\bibnamefont {Sanguinetti}}, \ and\ \bibinfo
  {author} {\bibfnamefont {R.~T.}\ \bibnamefont {Thew}},\ }\href {\doibase
  10.1364/oe.22.017246} {\bibfield  {journal} {\bibinfo  {journal} {Optics
  express}\ }\textbf {\bibinfo {volume} {22}},\ \bibinfo {pages} {17246}
  (\bibinfo {year} {2014})},\ \Eprint {http://arxiv.org/abs/1403.6740}
  {1403.6740} \BibitemShut {NoStop}%
\bibitem [{\citenamefont {Clausen}\ \emph {et~al.}(2014)\citenamefont
  {Clausen}, \citenamefont {BussiÃšres}, \citenamefont {Tiranov},
  \citenamefont {Herrmann}, \citenamefont {Silberhorn}, \citenamefont {Sohler},
  \citenamefont {Afzelius},\ and\ \citenamefont {Gisin}}]{Clausen2014}%
  \BibitemOpen
  \bibfield  {author} {\bibinfo {author} {\bibfnamefont {C.}~\bibnamefont
  {Clausen}}, \bibinfo {author} {\bibfnamefont {F.}~\bibnamefont
  {BussiÃšres}}, \bibinfo {author} {\bibfnamefont {A.}~\bibnamefont
  {Tiranov}}, \bibinfo {author} {\bibfnamefont {H.}~\bibnamefont {Herrmann}},
  \bibinfo {author} {\bibfnamefont {C.}~\bibnamefont {Silberhorn}}, \bibinfo
  {author} {\bibfnamefont {W.}~\bibnamefont {Sohler}}, \bibinfo {author}
  {\bibfnamefont {M.}~\bibnamefont {Afzelius}}, \ and\ \bibinfo {author}
  {\bibfnamefont {N.}~\bibnamefont {Gisin}},\ }\href {\doibase
  10.1088/1367-2630/16/9/093058} {\bibfield  {journal} {\bibinfo  {journal}
  {New Journal of Physics}\ }\textbf {\bibinfo {volume} {16}},\ \bibinfo
  {pages} {093058} (\bibinfo {year} {2014})},\ \Eprint
  {http://arxiv.org/abs/1405.6486} {1405.6486} \BibitemShut {NoStop}%
\bibitem [{\citenamefont {Franson}(1989)}]{Franson1989}%
  \BibitemOpen
  \bibfield  {author} {\bibinfo {author} {\bibfnamefont {J.~D.}\ \bibnamefont
  {Franson}},\ }\href {\doibase 10.1103/physrevlett.62.2205} {\bibfield
  {journal} {\bibinfo  {journal} {Physical Review Letters}\ }\textbf {\bibinfo
  {volume} {62}},\ \bibinfo {pages} {2205} (\bibinfo {year}
  {1989})}\BibitemShut {NoStop}%
\bibitem [{\citenamefont {Zielnicki}\ \emph {et~al.}(2018)\citenamefont
  {Zielnicki}, \citenamefont {Garay-Palmett}, \citenamefont {Cruz-Delgado},
  \citenamefont {Cruz-Ramirez}, \citenamefont {O’Boyle}, \citenamefont
  {Fang}, \citenamefont {Lorenz}, \citenamefont {U’Ren},\ and\ \citenamefont
  {Kwiat}}]{Zielnicki2018}%
  \BibitemOpen
  \bibfield  {author} {\bibinfo {author} {\bibfnamefont {K.}~\bibnamefont
  {Zielnicki}}, \bibinfo {author} {\bibfnamefont {K.}~\bibnamefont
  {Garay-Palmett}}, \bibinfo {author} {\bibfnamefont {D.}~\bibnamefont
  {Cruz-Delgado}}, \bibinfo {author} {\bibfnamefont {H.}~\bibnamefont
  {Cruz-Ramirez}}, \bibinfo {author} {\bibfnamefont {M.~F.}\ \bibnamefont
  {O’Boyle}}, \bibinfo {author} {\bibfnamefont {B.}~\bibnamefont {Fang}},
  \bibinfo {author} {\bibfnamefont {V.~O.}\ \bibnamefont {Lorenz}}, \bibinfo
  {author} {\bibfnamefont {A.~B.}\ \bibnamefont {U’Ren}}, \ and\ \bibinfo
  {author} {\bibfnamefont {P.~G.}\ \bibnamefont {Kwiat}},\ }\href {\doibase
  10.1080/09500340.2018.1437228} {\bibfield  {journal} {\bibinfo  {journal}
  {Journal of Modern Optics}\ }\textbf {\bibinfo {volume} {65}},\ \bibinfo
  {pages} {1141} (\bibinfo {year} {2018})},\ \Eprint
  {http://arxiv.org/abs/1801.01195} {1801.01195} \BibitemShut {NoStop}%
\bibitem [{\citenamefont {Luo}\ \emph {et~al.}(2015)\citenamefont {Luo},
  \citenamefont {Herrmann}, \citenamefont {Krapick}, \citenamefont {Brecht},
  \citenamefont {Ricken}, \citenamefont {Quiring}, \citenamefont {Suche},
  \citenamefont {Sohler},\ and\ \citenamefont {Silberhorn}}]{Luo_2015}%
  \BibitemOpen
  \bibfield  {author} {\bibinfo {author} {\bibfnamefont {K.-H.}\ \bibnamefont
  {Luo}}, \bibinfo {author} {\bibfnamefont {H.}~\bibnamefont {Herrmann}},
  \bibinfo {author} {\bibfnamefont {S.}~\bibnamefont {Krapick}}, \bibinfo
  {author} {\bibfnamefont {B.}~\bibnamefont {Brecht}}, \bibinfo {author}
  {\bibfnamefont {R.}~\bibnamefont {Ricken}}, \bibinfo {author} {\bibfnamefont
  {V.}~\bibnamefont {Quiring}}, \bibinfo {author} {\bibfnamefont
  {H.}~\bibnamefont {Suche}}, \bibinfo {author} {\bibfnamefont
  {W.}~\bibnamefont {Sohler}}, \ and\ \bibinfo {author} {\bibfnamefont
  {C.}~\bibnamefont {Silberhorn}},\ }\href {\doibase
  10.1088/1367-2630/17/7/073039} {\bibfield  {journal} {\bibinfo  {journal}
  {New Journal of Physics}\ }\textbf {\bibinfo {volume} {17}},\ \bibinfo
  {pages} {073039} (\bibinfo {year} {2015})}\BibitemShut {NoStop}%
\bibitem [{\citenamefont {Luo}\ \emph {et~al.}(2017)\citenamefont {Luo},
  \citenamefont {Herrmann},\ and\ \citenamefont {Silberhorn}}]{Luo_2017}%
  \BibitemOpen
  \bibfield  {author} {\bibinfo {author} {\bibfnamefont {K.-H.}\ \bibnamefont
  {Luo}}, \bibinfo {author} {\bibfnamefont {H.}~\bibnamefont {Herrmann}}, \
  and\ \bibinfo {author} {\bibfnamefont {C.}~\bibnamefont {Silberhorn}},\
  }\href {\doibase 10.1088/2058-9565/aa6b8e} {\bibfield  {journal} {\bibinfo
  {journal} {Quantum Science and Technology}\ }\textbf {\bibinfo {volume}
  {2}},\ \bibinfo {pages} {024002} (\bibinfo {year} {2017})}\BibitemShut
  {NoStop}%
\end{thebibliography}%
\end{document}